\documentclass[conference]{IEEEtran}
\IEEEoverridecommandlockouts
\usepackage{cite}
\usepackage{amsmath,amssymb,amsfonts}
\usepackage{graphicx}
\usepackage{textcomp}
\usepackage{xcolor}
\usepackage{threeparttable}
\usepackage{algorithm}  
\usepackage{algorithmicx}
\usepackage{algpseudocode}
\usepackage{graphicx}
\usepackage{textcomp}
\usepackage{pdflscape}
\usepackage{booktabs}
\usepackage{multirow}
\usepackage{stfloats}
\usepackage[normalem]{ulem}
\useunder{\uline}{\ul}{}
\def\BibTeX{{\rm B\kern-.05em{\sc i\kern-.025em b}\kern-.08em
		T\kern-.1667em\lower.7ex\hbox{E}\kern-.125emX}}
\begin{document}
	
	\title{Weight-based Channel-model Matrix Framework provides a reasonable solution for EEG-based cross-dataset emotion recognition\\
		\thanks{
			\textsuperscript{\rm 1} Gansu Provincial Key Laboratory of Wearable Computing, School of Information Science and Engineering, Lanzhou University, Lanzhou, China.\\
			\textsuperscript{\rm 2} Brain Health Engineering Laboratory, School of Medical Technology, Beijing Institute of Technology, Beijing, China.\\
			\textsuperscript{\rm 3} School of Information Science and Engineering, Inner Mongolia University, Inner Mongolia, China.\\
			\textsuperscript{\rm 4} Shandong Academy of Intelligent Computing Technology, Jinan, China.\\
			\textsuperscript{\rm 5} Chinese Academy of Sciences Center for Excellence in Brain Science and Intelligence Technology, Shanghai Institutes for Biological Sciences, Chinese Academy of Sciences, Shanghai, China.\\
			\textsuperscript{\rm 6} Joint Research Center for Cognitive Neurosensor Technology of Lanzhou University and Institute of Semiconductors, Chinese Academy of Sciences, Lanzhou, China.\\
			\textsuperscript{\rm 7} Open Source Software and Real-Time System, Lanzhou University, Ministry of Education, Lanzhou, China.\\
			\textsuperscript{\rm *} Corresponding author. \\
			Email address:\{chenhy2021, 220220942890, zhujing, sunsht17, lijx18, shaoxx19, lijunxiang19, lixwei, bh\}@lzu.edu.cn.
		}
	}
	
	\author {
		\textbf{
		Huayu Chen\textsuperscript{\rm 1}, 
		Huanhuan He\textsuperscript{\rm 1},
		Jing Zhu\textsuperscript{\rm 1},
		Shuting Sun\textsuperscript{\rm 2},
		Jianxiu Li\textsuperscript{\rm 3},
		Xuexiao Shao\textsuperscript{\rm 1},
		Junxiang Li\textsuperscript{\rm 1},}
		\\
		\textbf{
		Xiaowei Li\textsuperscript{\rm 1,4*},
		Bin Hu\textsuperscript{\rm 1,2,5,6,7*}}
	}

	\maketitle
	
	\begin{abstract}
	Cross-dataset emotion recognition as an extremely challenging task in the field of EEG-based affective computing is influenced by many factors, which makes the universal models yield unsatisfactory results. Facing the situation that lacks EEG information decoding research, we first analyzed the impact of different EEG information(individual, session, emotion and trial) for emotion recognition by sample space visualization, sample aggregation phenomena quantification, and energy pattern analysis on five public datasets. Based on these phenomena and patterns, we provided the processing methods and interpretable work of various EEG differences. Through the analysis of emotional feature distribution patterns, the Individual Emotional Feature Distribution Difference(IEFDD) was found, which was also considered as the main factor of the stability for emotion recognition. After analyzing the limitations of traditional modeling approach suffering from IEFDD, the Weight-based Channel-model Matrix Framework(WCMF) was proposed. To reasonably characterize emotional feature distribution patterns, four weight extraction methods were designed, and the optimal was the correction T-test(CT) weight extraction method. Finally, the performance of WCMF was validated on cross-dataset tasks in two kinds of experiments that simulated different practical scenarios, and the results showed that WCMF had more stable and better emotion recognition ability.
	\end{abstract}
	
	
	\section{Introduction}
	The affective brain-computer interface\cite{huang2019eeg} is an important component of brain-computer interface, which has suffered from the severe impact by EEG individual information, causing the universal model is unable to give stable and effective performance on cross-session\cite{wei2020eeg,peng2021self,chen2021ms}, cross-subject\cite{fdez2021cross,shen2022contrastive,huang2022generator}, cross-dataset\cite{cimtay2020investigating,lan2018domain,he2022adversarial}, and other emotion classification tasks. In previous works, many studies demonstrated that machine learning algorithms could learn the emotional difference information delivered by EEG signals for single subject\cite{yin2021eeg,huang2021differences}. However, due to the existence of EEG individual differences, it is difficult to find common emotion knowledge among different subjects to build universal emotion recognition models. Thus, the universal model usually produces extremely unstable results on unknown subjects and unknown sessions, and it is more challenging to implement stable emotion recognition between datasets. Importantly, only if affective brain-computer interfaces that could provide stable and effective performance on cross-session, cross-subject and cross-dataset tasks, it could be expected to deal with complex and variable real-life emotion recognition scenarios.
	
	In the cross-session emotion recognition classification task, many studies have shown that there are some differences between the EEG data collected from single subject in different time periods, which makes subject-dependent models have inconsistent performance on session datasets, and the abrupt drop often occurs in classification accuracy. He et al. performed the intra-day task and cross-day task on a 12-person dataset separately. The results showed that the subject-dependent model had an excellent performance on the intra-day task, but it did suffer from the accuracy drop on cross-day task. They tried to alleviate the problem of session sample domain differences by the transfer component analysis algorithm(TCA), which obtained some improvement\cite{he2022cross}. Meanwhile, Lin et al. proposed a robust principal component analysis(RPCA)-embedded transfer learning(TL) to obviate intra- and inter-individual differences\cite{lin2019constructing}. In the Multiple Day Music-listening experiment using Emotiv(MDME) dataset that included 12 subjects for the cross-session task, the PCA sample space visualization revealed that the session sample domains of different subjects had different degrees of variation, and the RPCA had the ability to close the distance between session domains, which resulted in alleviating the session differences. This cross-time EEG difference was also mentioned in other studies\cite{bao2021two,wei2020eeg,peng2021self}. Therefore, the solution of cross-time EEG differences determines the performance of subject-dependent model in a significant way.
	
	For the cross-subject emotion recognition task, the impact of individual differences is more powerful, where the universal model often exhibits poor performance when facing unknown subjects\cite{cai2022cross}. To overcome the problem that emotion recognition model had poor generalization across subjects, Meng et al. proposed Deep Subdomain Associate Adaptation Network(DSAAN) to build a source-to-target domain transfer network by minimizing the sum of source-domain classification loss and Subdomain Associate Loop(SAL), and the result of cross-subject positive-negative emotion classification experiment in SEED was 89.23 ± 1.93\%\cite{meng2022deep}. Zhao et al. learned the group public component and subject private component by shared encoder and private encoder respectively, this framework achieved domain transfer by feature encoding. After private component extraction for unknown subjects, the more similar subject private classifiers would be granted higher weights, which would influence the joint classification of private and shared classifiers, and the result of 86.7 ± 7.1\% was obtained on the SEED dataset\cite{zhao2021plug}. In general, the main solutions of domain differences caused by individual information are transfer learning\cite{huang2022generator,luo2018wgan,li2022cross,wang2021deep} and deep learning\cite{li2021cross,bao2021two,he2022adversarial}. For the results, both methods can obtain excellent results, but both miss the analysis of EEG information and the interpretability of EEG individual differences, which as the basic theory of affective brain-computer interfaces is more important for EEG-based emotion recognition. In addition, the ideal results on one dataset still had limitations.
	
	Compared with the cross-subject emotion recognition task, the cross-dataset task not only needs to deal with EEG individual differences, but also needs to consider differences in acquisition devices, emotion stimulation materials and emotion modeling approaches. Therefore, the cross-dataset emotion classification task is the most challenging. If a reasonable and effective solution could be given, it would substantially improve the performance of affective brain-computer interface. Lan et al. performed positive-neutral-negative emotion classification experiments on SEED\cite{zheng2015investigating,duan2013differential} and DEAP\cite{koelstra2011deap} datasets with various domain-adaptive algorithms, and there were significant differences between two datasets in terms of acquisition devices, emotion stimulation materials and emotion modeling approaches. In the cross-dataset experiments, although the domain adaptation algorithm could make some improvements compared to the baseline, the poor results were obtained with accuracies ranging from 30\% to 50\%\cite{lan2018domain}. Ni et al. proposed new domain adaptation sparse representation classifier(DASRC) applied to the cross-domain emotion recognition task, and cross-subject tasks results were similar to other research in both the DEAP and SEED, but the accuracy drop still occured in the cross-dataset task, which fell to 50\% directly\cite{ni2021domain}. Similarly, this situation occurred in other research about cross-dataset emotion recognition, which indicated that some unknown factors contributed to universal model usually show the unsatisfactory emotion recognition ability on the cross-dataset task\cite{tao2021multi,pandey2021subject,kuang2021cross}.
	
	In summary, complex EEG individual information has a huge impact on different emotion recognition tasks. Although these differences are well accepted in the EEG-based affective computing field and there exists solutions to deal with individual differences, the in-depth investigation about EEG information is rare, which results in deep learning and transfer learning methods have certain effects but still lack interpretable theoretical support about EEG signals. Therefore, it is necessary to clear up the components of EEG information for affective computing. Only if we know the impact of different information on universal models, it is possible to draw up the processing methods for different EEG information, and attempt to achieve more stable affective brain-computer interfaces. Based on the above ideas, we conducted an interpretability study of EEG information, analyzed the impacts of different information on EEG sample distribution, and designed solutions to cope with difficult cross-dataset tasks based on the obtained knowledge of different EEG information. The major contributions are as follows.
	
	(1) Through the analysis of feature extraction algorithm, we compared the performance and computational principles of differential entropy and fractal dimension, analyzed the meanings of two feature for EEG signals from the view of energy, and simplified the calculation process.
	
	(2) For the complex EEG information, the different sample aggregation phenomena were observed by sample space visualization based on power feature. And the difference information ratio and sample aggregation degree were quantified for each dataset, all dataset were the mixture of different EEG information, which influence the distribution of samples. Based on the EEG energy distribution patterns, we analyzed the influence of different information and concluded the cause of different sample aggregation phenomena. Finally, a reasonable explanation work for different sample aggregation phenomena was presented.
	
	(3) For individual emotional information, the individual emotional feature distribution patterns were obtained by channel emotion classification, the individual emotional feature distribution difference were found, and its impact on the construction of universal model was analyzed. In order to deal with this difference appropriately, we designed different weight extraction methods to quantify emotional feature distribution patterns, and proposed the weight-based channel-model matrix framework, which made the universal model more stable and effective on cross-dataset tasks.
	
	This paper is organized as follows. Section II reviews the five datasets we used in this paper, and documents the data preprocessing process, including signal preprocessing, feature extraction, and sample selection. Section III analyzes and quantifies the impact of different EEG information on sample distribution, and gives an explanation of EEG sample aggregation phenomena based on energy distribution patterns. Section IV explains the advantages of channel-model matrix modeling over traditional multidimensional modeling based on emotional feature distribution patterns, and introduces the design process of four weight extraction methods and trial correction strategy. Section V describes the details and results of two experimental protocols. Scetion VI is the discussions about the weight-based channel-model matrix framework and the EEG-based cross-dataset emotion recognition.

	\section{dataset preprocessing}
	
	\subsection{Datasets}
	Many stuides have attempted to perform cross-dataset tasks on SEED and DEAP datasets. But due to the differences in EEG acquisition devices, data specifications, emotion stimulation materials and emotion label modeling between the two datasets, the cross-dataset classification task is extremely challenging. Therefore, we selected five similar public datasets: SEED, SEED-IV\cite{zheng2018emotionmeter}, SEED-V\cite{liu2021comparing}, RCLS\cite{li2019eeg} and MPED\cite{song2019mped}. The commonalities of these datasets are as follows.
	
	(1) The EEG acquisition devices are the 62-channel ESI NeuroScan System with international 10-20 system of EEG caps.
	
	(2) Paradigms all use video clips to induce emotions, and all datasets contain positive, negative, and neutral emotions.
	
	(3) The category labels are discrete emotion modeling method rather than valence-arousal emotion modeling method.
	
	\subsection{Signal Preprocessing process}
	Signal preprocessing plays an important role in EEG signal application, which clean the EEG signal through a series of processing steps. Firstly, we used the AutoMagic\cite{pedroni2019automagic} automatic preprocessing framework to perform the following operations on all EEG signals: (1)Band-pass frequency filter between 1Hz and 50Hz. (2)Taking Fpz electrode signals as reference, EOG regression was employed to eliminate EOG\cite{croft2000eog,parra2005recipes}. (3)IClabel was used to eliminate artifacts\cite{pion2019iclabel}. (4)Bad channel signal interpolation and signal quality assessment based on some signal-level metrics. (5)The average re-referencing was performed and signals were downsampled to 200Hz. (6)All signals were segmented to 4s. Considering the differences in the paradigm duration of different datasets, different numbers of samples are generated from different datasets.
	
	\subsection{Feature extraction}
	Based on the feature analysis of differential entropy and fractal dimension in Appendix A, we knew how both features work in the EEG signal measurement, and we extracted the simplified features: variance(differential entropy\cite{duan2013differential}) and power(Katz’s fractal dimension\cite{katz1988fractals,jacob2019application}). The calculation formulas are as follows, where $x$ represents the EEG signal. The detailed analysis can be found in Appendix A.
	
	\begin{equation}
	variance(x) = log(\frac{\sum_{1}^{n} (x_{i} -\bar{x} )^{2} }{n})
	\end{equation}
		
	\begin{equation}
	power(x)=log(\sum_{1}^{n-1} (x_{i} -x_{i+1})^{2})
	\end{equation}
	
	\begin{equation}
	Inverse\_power(x)=\frac{1}{1+power(x)} 
	\end{equation}
	
	\subsection{Sample selection}
	Due to the varying paradigm lengths between datasets, there were differences in the number of segmented samples. We would like to build balanced emotion datasets with a certain number of data according to the sample number of different datasets. The samples were selected based on the following two principles.
	
	(1) For emotion analysis, the emotions(positive, negative, neutral) are selected for analysis, which of them are the common emotions of each dataset.
	
	(2) To ensure the balance of the dataset, n samples are selected randomly on each trial's segmented dataset in different datasets.
	
	The specification of datasets after selection is as follows.
	
	\begin{table}[h]
		\centering
		\caption{SPECIFICATIONS FOR DIFFERENT DATASETS}
		\label{tab1}
		\resizebox{\linewidth}{!}{%
		\begin{threeparttable}
			\begin{tabular}{cccccl}
				\hline
				\multicolumn{1}{l}{\textbf{Dataset}} &
				\multicolumn{1}{l}{\textbf{Subject}} &
				\multicolumn{1}{l}{\textbf{Session}} &
				\multicolumn{1}{l}{\textbf{Trial}} &
				\multicolumn{1}{l}{\textbf{Sample}} &
				\textbf{Total} \\ \hline
				\textbf{SEED}    & 15 & 3 & 15 & 50 & 33750 \\
				\textbf{SEED-IV} & 15 & 3 & 18 & 10 & 8100  \\
				\textbf{SEED-V}  & 16 & 3 & 9 & 20 & 8640\\
				\textbf{RCLS}    & 14 & 1 & 15 & 20 & 4200  \\
				\textbf{MPED}    & 22 & 2 & 6  & 40 & 10560 \\ \hline
			\end{tabular}%
			\small Note: Trial: the number of trial we chose. Sample: the number of samples for each trial. Total: the total number of sample for each dataset.
		\end{threeparttable}
	}
	\end{table}
	
	\section{Investigation of EEG information}
	
	\subsection{Visualization of the sample space}
	For different datasets, the visualization of sample distribution were obtained by t-SNE\cite{maaten2008visualizing} in intra-session and multi-sessions scenarios respectively. After observation, we found that there were four kinds of sample distribution patterns under variance and power features: emotional aggregation, trial aggregation, session aggregation, and individual aggregation.
	
	\subsubsection{Emotional aggregation and trial aggregation in intra-session scenes}

\begin{figure}[htb]
		\centerline{\includegraphics[width=\linewidth]{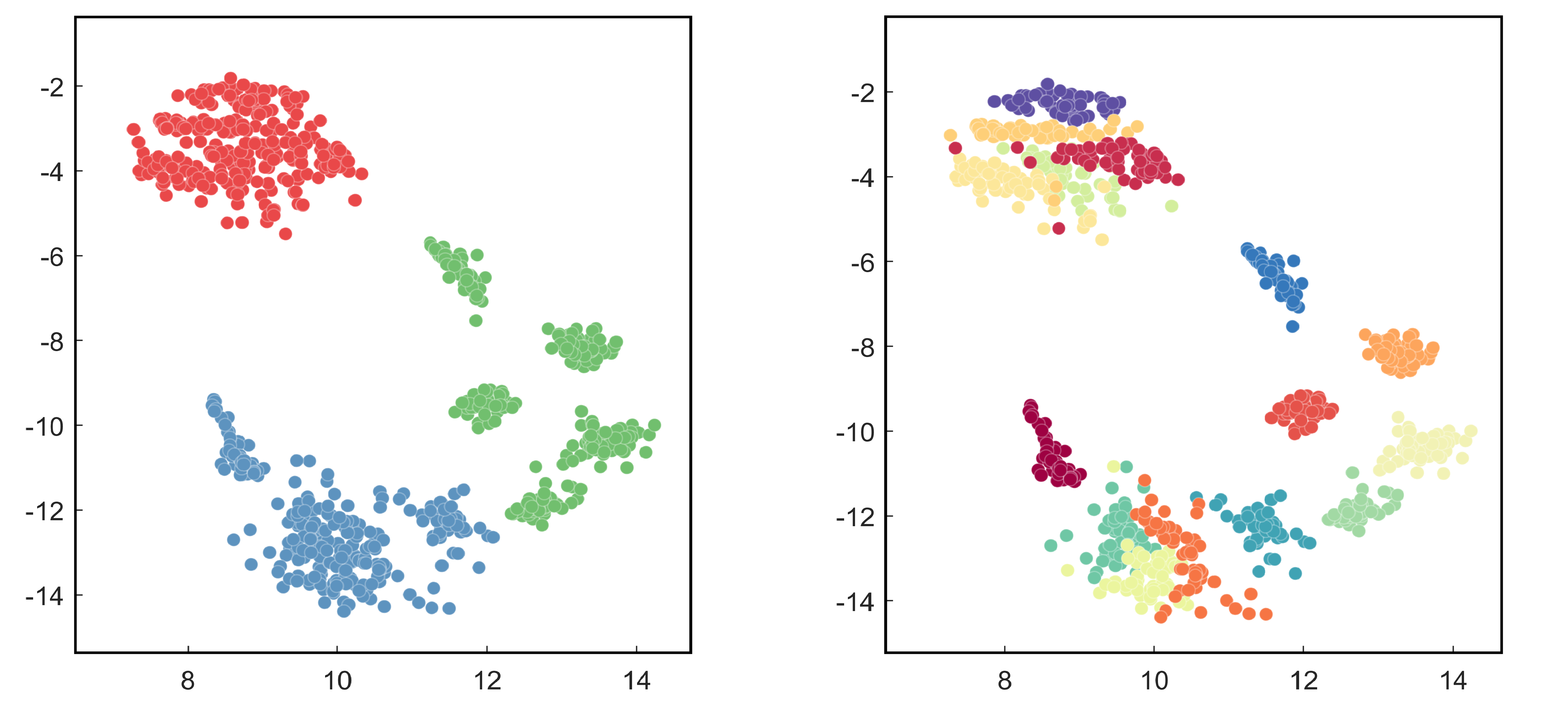}}
		\caption{The sample space visualization in the intra-session scenario under power feature. The left subplot is the sample space visualization marked by emotions(positive-red, negative-blue, neutral-green), and the right subplot is the sample space visualization marked by trials.}
		\label{fig1}
	\end{figure}

	Firstly, we visualized each session dataset and analyzed the distribution of emotional samples under different session data. The details are shown in Fig. \ref{fig1}.
	
	 It was found that there existed different degrees of emotional sample aggregation under different data. In particular, on the basis of emotional sample aggregation, there are several samples clustered into small clusters among the emotional clusters. We assumed that the samples were not only aggregated by emotion, but also existed trial aggregation phenomenon. Therefore, we marked the different trial samples with different colors, and trial aggregation phenomenon were found.

	\subsubsection{Session aggregation and individual aggregation in multi-sessions scenarios}
	
	Considering that individual difference is a widely-accepted attribute of EEG, we first investigated the relationship between different session data for each subject. For the multi-session dataset, we visualized the sample distribution of multiple session data belonging to a single subject and observed the distribution pattern between different session data. The visualization results are as follows.
	
	\begin{figure}[htb]
		\centerline{\includegraphics[width=\linewidth]{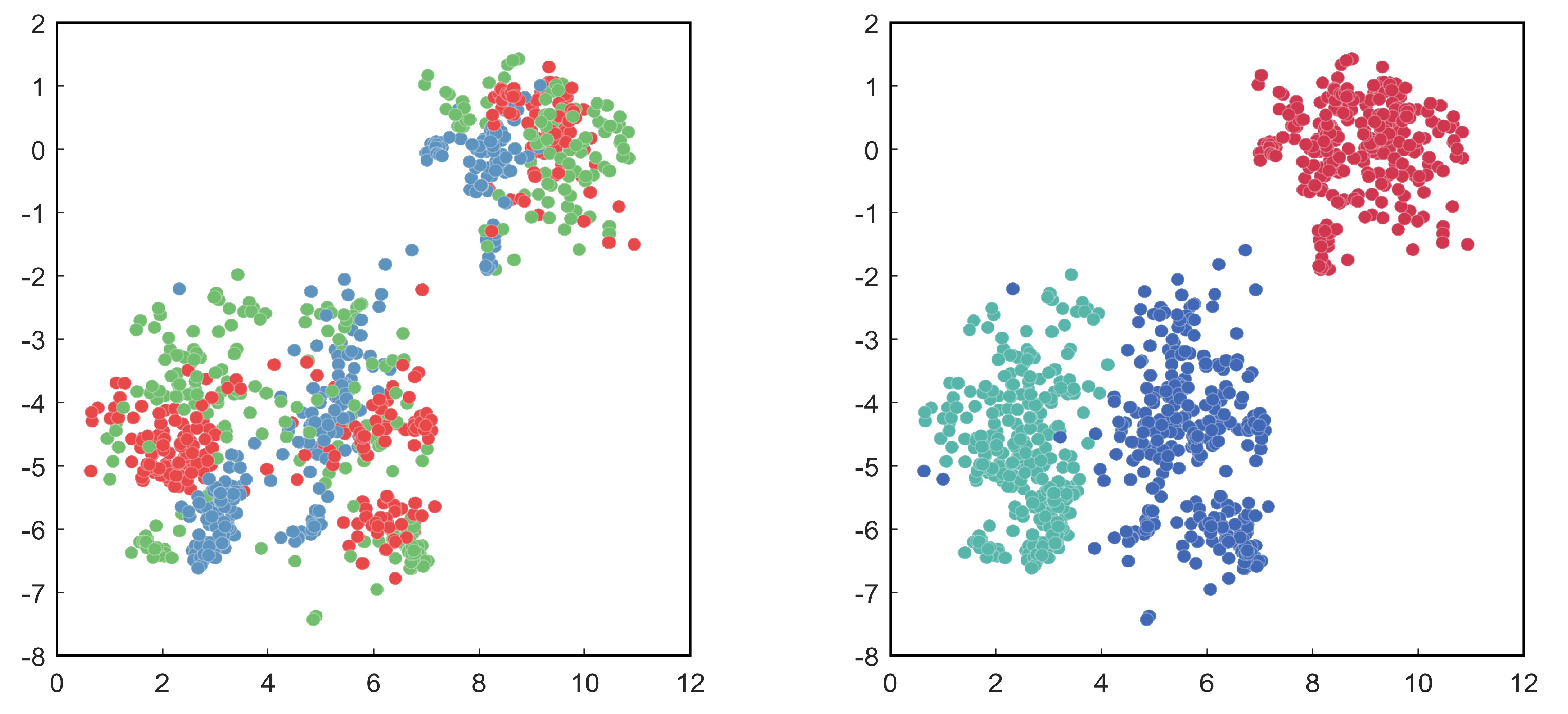}}
		\caption{The sample space visualization in the single-subject scenario under power feature, the left subplot is the sample space visualization marked by emotions(positive-red, negative-blue, neutral-green), and the right subplot is the sample space visualization marked by sessions.}
		\label{fig2}
	\end{figure}
	
	As shown in Fig. \ref{fig2}, the samples are clustered by session. There are different degrees of emotional aggregation in different session domains. This phenomenon indicates the following two points: 
	
	(1) The subjects' brain states are stable during each session, which lead to a tendency of aggregation on each session samples. 
	
	(2) Even though different sessions generated from the same subject, there also existed differences between different sessions.

\begin{figure}[htb]
	\centerline{\includegraphics[width=\linewidth]{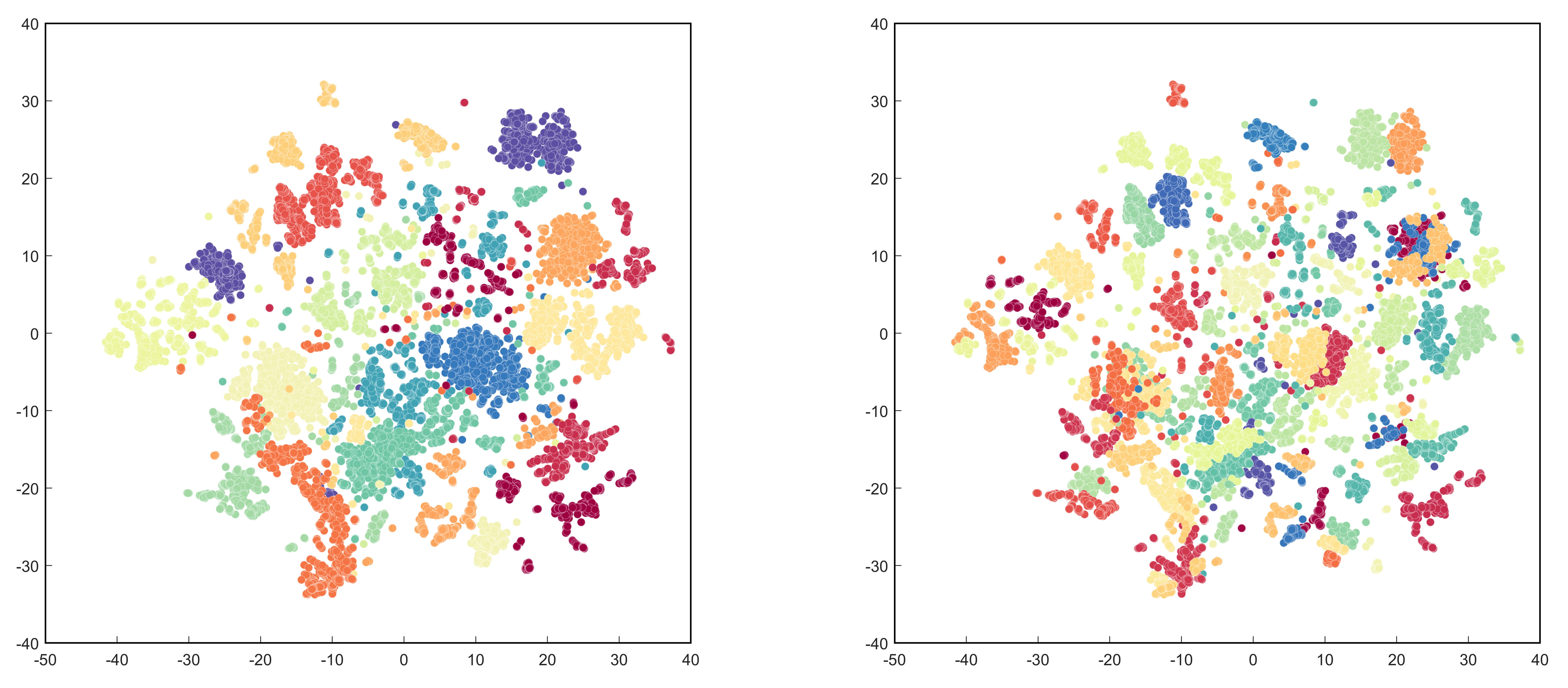}}
	\caption{The sample space visualization in the multi-subjects scenario under the power feature, the left subplot is the sample space visualization marked by subjects, and the right subplot is the sample space visualization marked by sessions.}
	\label{fig3}
	\end{figure}

	After observing the session aggregation phenomenon in the intra-subject scenario, it is necessary to investigate the influence of session aggregation phenomenon and the relationship between sessions of different subjects in multi-subjects scenarios. The visualizations of the multi-subjects sample distribution are shown in Fig. \ref{fig3}.

	As shown in the left subplot of Fig. \ref{fig3}, samples marked by subject, there were large differences between the samples belonging to different subjects. Due to the impact of individual differences in EEG, sample space shows the individual aggregation phenomenon. As shown in the right subplot of Fig. \ref{fig3}, samples marked by session, we found that the sample distribution in multi-subjects scenario still retained the session aggregation phenomenon, while the samples of different session presented different degrees of aggregation. In most cases, each individual cluster is composed of session sub-clusters.
	
	In conclusion, the samples present a tendency of session aggregation on the basis of individual aggregation in the multi-subjects scenario, which reveals that the EEG individual information contains session difference information. However, these individual information are extremely unhelpful to the construction of universal model. In intra-session scenario, the samples show trial aggregation on the base of emotional aggregation, which reveal that the emotional difference information includes the trial difference information. Therefore, the EEG information can be decomposed into four kinds of information(individual, session, emotion, trial), each of which causes different aggregation phenomena. The relationship of sample clusters under four aggregation phenomena is individual $\supseteq$ session $\supseteq$ emotion $\supseteq$ trial.
	
	~\\
	\subsection{Quantification of sample aggregation phenomena}
	
	To further know the degree of different aggregation phenomena for each dataset, two metrics, the ratio of difference features and the sample aggregation coefficient(SAC), were used to quantify various aggregation phenomena.
	
	The first metric is the ratio of difference features. Taking the dataset with n-dimensional features as an example, Pearson correlation analysis is employed between each dimensional data and the label of corresponding difference, and the features with P$\le$0.05 are considered as difference features, the difference information of which are the main factors of sample aggregation phenomena. Finally, the ratio between the number of difference features and the total number of features is calculated to measure the difference information component of the dataset\cite{chen2021personal}.
	
	Secondly, we combined the t-sne and k-means algorithms for the calculation of the sample aggregation coefficient, which represented the degree of intra-cluster sample aggreagation influenced by different EEG information. The details are shown in Appendix B.

	Under the variance feature and inverse power feature, two metrics were calculated. The results are as follows.
	
	\clearpage
	\begin{table*}[htb]
		\centering
		\caption{THE RESULTS OF THE DIFFERENT DATASET UNDER TWO INDICATORS}
		\label{tab2}
		\resizebox{0.9\linewidth}{!}{%
			\begin{tabular}{clcccccccc}
				\hline
				\multicolumn{1}{l}{\multirow{2}{*}{\textbf{Feature}}} &
				\multirow{2}{*}{\textbf{DataSet}} &
				\multicolumn{4}{c}{\textbf{the ratio of difference feature}} &
				\multicolumn{4}{c}{\textbf{Sample Aggregation Coefficient}} \\
				\multicolumn{1}{l}{} &
				&
				\textbf{Emotion} &
				\textbf{Trial} &
				\textbf{Session} &
				\textbf{Subject} &
				\textbf{Emotion} &
				\textbf{Trial} &
				\textbf{Session} &
				\textbf{Subject} \\ \hline
				\multirow{5}{*}{\textbf{\begin{tabular}[c]{@{}c@{}}Inverse\\ Power\end{tabular}}} &
				\textbf{SEED} &
				56.98\% &
				74.83\% &
				98.36\% &
				\multicolumn{1}{c|}{100.00\%} &
				79.78\% &
				89.75\% &
				77.42\% &
				77.60\% \\
				& \textbf{SEED-IV} & 59.45\% & 58.72\% & 67.21\% & \multicolumn{1}{c|}{77.05\%} & 81.52\% & 81.79\% & 90.97\% & 73.75\% \\
				& \textbf{SEED-V}  & 57.65\% & 60.18\% & 93.44\% & \multicolumn{1}{c|}{93.44\%} & 77.42\% & 91.62\% & 79.06\% & 80.98\% \\
				& \textbf{MPED}    & 78.02\% & 70.72\% & 88.52\% & \multicolumn{1}{c|}{88.52\%} & 84.01\% & 97.07\% & 79.13\% & 73.76\% \\
				& \textbf{RCLS}    & 76.93\% & 72.25\% & 90.16\% & \multicolumn{1}{c|}{90.16\%} & 86.08\% & 95.74\% & 97.54\% & 97.54\% \\ \hline
				\multirow{5}{*}{\textbf{Variance}} &
				\textbf{SEED} &
				57.41\% &
				14.32\% &
				100.00\% &
				\multicolumn{1}{c|}{100.00\%} &
				79.55\% &
				72.75\% &
				76.81\% &
				64.57\% \\
				& \textbf{SEED-IV} & 44.85\% & 52.09\% & 90.16\% & \multicolumn{1}{c|}{90.16\%} & 80.77\% & 68.63\% & 90.71\% & 63.67\% \\
				& \textbf{SEED-V}  & 48.46\% & 52.63\% & 93.44\% & \multicolumn{1}{c|}{91.80\%} & 77.17\% & 80.64\% & 80.56\% & 79.05\% \\
				& \textbf{MPED}    & 66.54\% & 60.32\% & 91.80\% & \multicolumn{1}{c|}{91.80\%} & 87.06\% & 90.31\% & 87.31\% & 74.75\% \\
				& \textbf{RCLS}    & 53.98\% & 44.03\% & 77.05\% & \multicolumn{1}{c|}{77.05\%} & 79.99\% & 57.09\% & 74.86\% & 74.86\% \\ \hline
			\end{tabular}%
		}
	\end{table*}
	
	As shown in Table \ref{tab2}, the following conclusions were obtained:
	
	(1) For the ratio of difference features, there existed different numbers of features representing the differences of emotion, trial, session, and subject on all the datasets, which influenced the sample distribution.
	
	(2) Comparing the results of variance features and inverse power features under different datasets, we found that all datasets had similar results on the quantification of emotion, session and subject sample aggregation. However, the inverse power feature showed a higher ratio of difference features and a more stable trial aggregation phenomenon compared to the variance feature in the quantification of trial sample aggregation.
	
	~\\
	\subsection{Analysis of brain energy patterns}
	To further investigate the causes of these aggregation phenomena at the feature level, we draw energy distribution plot with power feature for each sample to analyze the differences from different views. And the power feature values of each sample were normalized to the range of 0 and 1. Therefore, the plots represent the relative energy distribution pattern.
	
	Firstly, we observed the energy patterns in different time periods of a single trial for trial aggregation phenomenon analysis, and it was found that the power feature had the property of cross-time stable patterns such as differential entropy\cite{zheng2015investigating}.
	
	\begin{figure}[h]
		\centerline{\includegraphics[width=\linewidth]{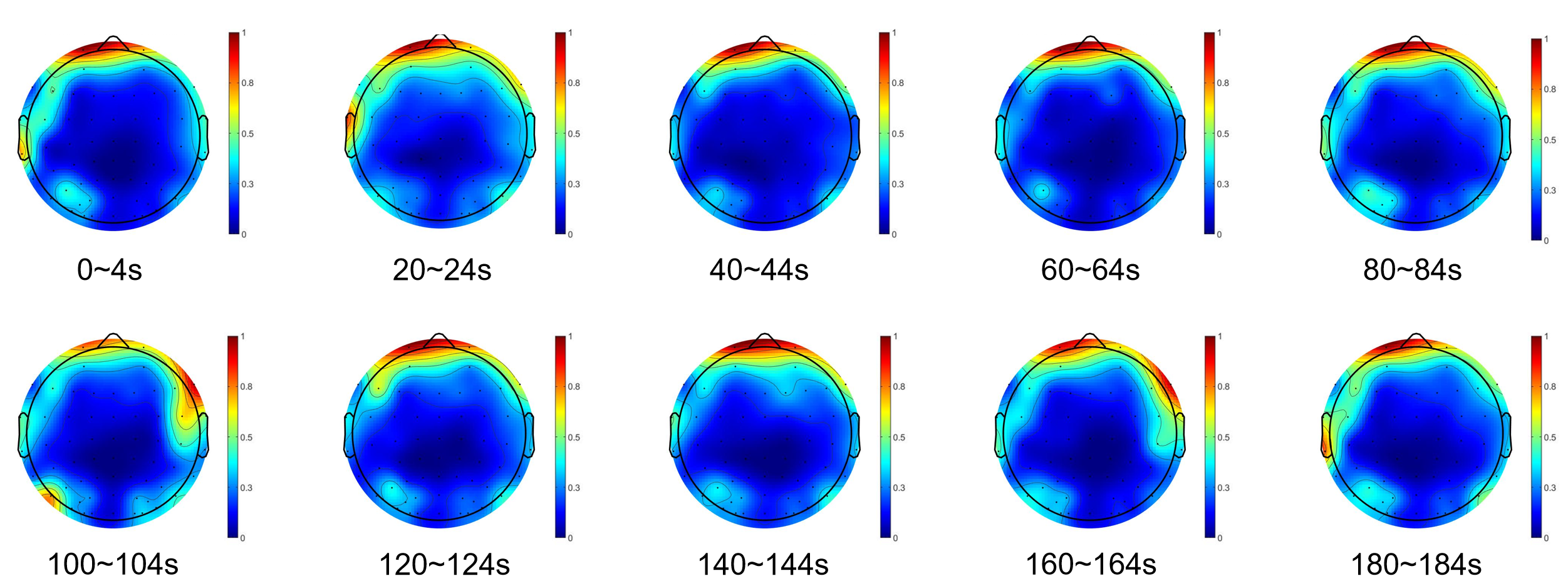}}
		\caption{Energy patterns under different time periods of a single trial.}
		\label{fig4}
	\end{figure}
	
	As shown in Fig. \ref{fig4}, although there were some differences in energy patterns between time periods, it could be found that brain energy patterns were stable during the whole trial. In addition, after observing the energy distribution of different subjects, we found that the areas with high energy of most subjects were mainly distributed in the frontal, temporal and occipital lobes. The energy in the parietal and anterior occipital lobes was lower than that in the above regions.
	
	Secondly, we visualized the energy patterns of different trials for each session, and compared the differences in energy patterns.
	~\\
	~\\
	\begin{figure}[h]
		\centerline{\includegraphics[width=\linewidth]{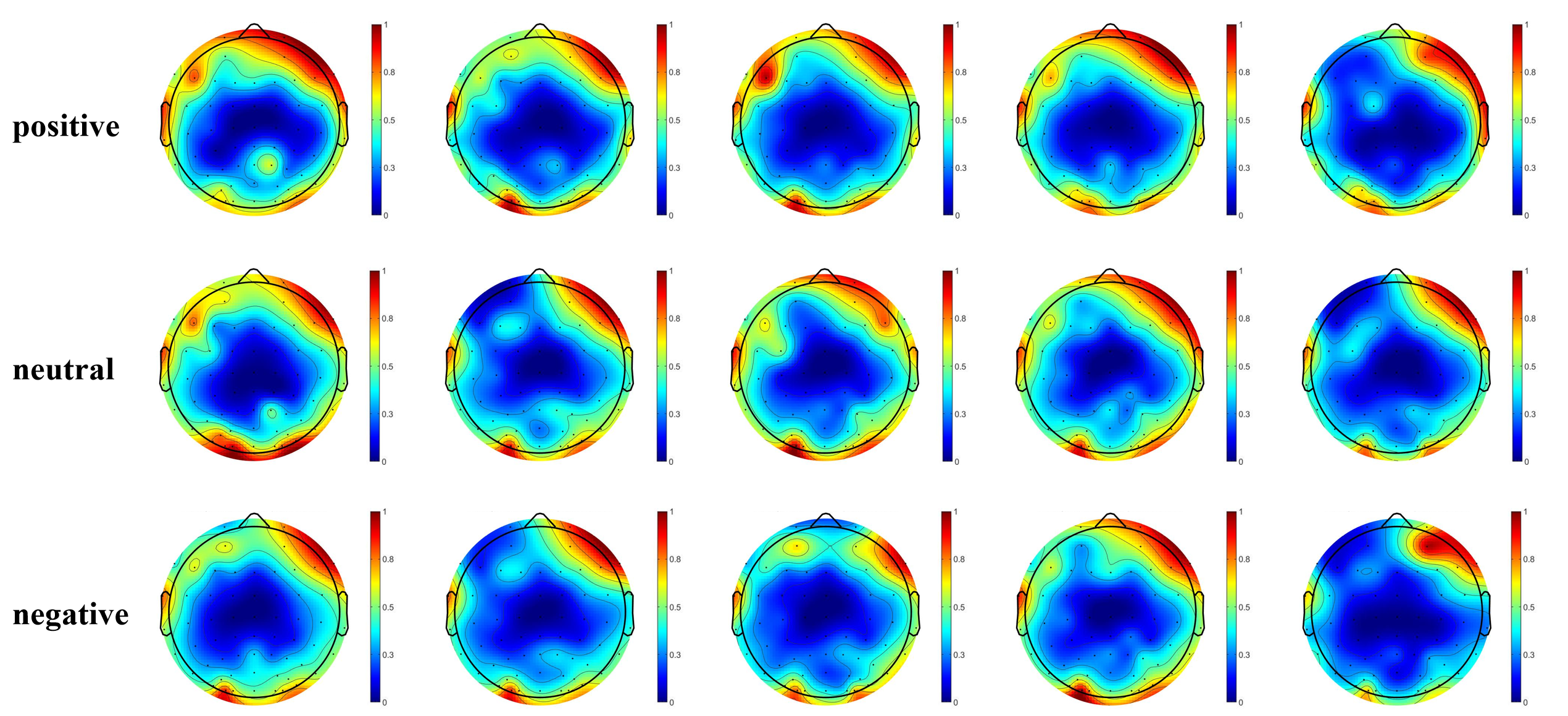}}
		\caption{Different trial energy patterns under a single session.}
		\label{fig5}
	\end{figure}

	As shown in Fig. \ref{fig5}, there are more or less differences in the trial energy patterns, which might be result from the energy change of brain activity under the stimulation of different trials during the same session. But trial energy distributions are still stable in general.
	
	Combining the above results, we assumed that the trial aggregation phenomenon was due to the similar energy patterns among trial samples. The gaps between trial clusters were influenced by the differences in trial energy patterns. In addition, the differences between trials were greater than intra-trial sample differences.
	
\begin{figure}[h]
		\centerline{\includegraphics[width=\linewidth]{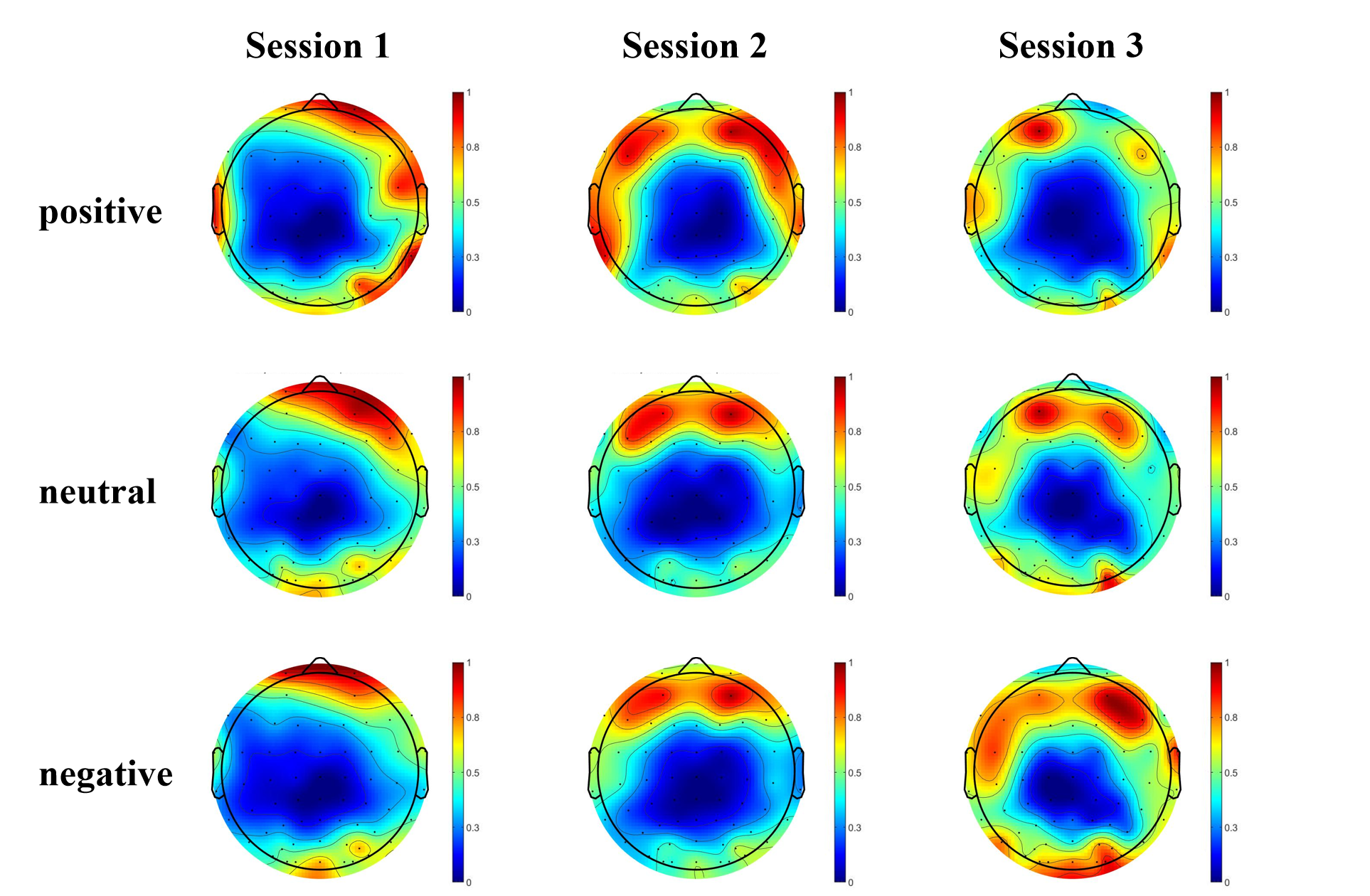}}
		\caption{The energy patterns of different emotions for a single subject under different sessions.}
		\label{fig6}
	\end{figure}

	After analysis of trial energy patterns, we combined trials’ energy pattern belonging to the same emotion for each session, and compared these energy patterns from the views of emotion and session. The energy pattern plots are shown in Fig. \ref{fig6}.
	
	\clearpage
	
	After comparison of different emotional energy patterns from a single session, it was found that the emotional differences were also caused by the differences in the energy distribution. In addition, the comparison of different sessions' energy patterns indicated that there existed greater energy pattern differences between sessions than that of emotion. Similarly, the phenomena of individual aggregation also resulted from greater energy differences.
	
	Combining the results of sample distribution visualization and energy pattern analysis, we knew that various sample aggregation phenomena were caused by different degrees of difference information, and the difference information mainly reflected the energy pattern differences.
	
	In general, we concluded the following 3 points to explain the difference in EEG information.
	
	(1) Emotional differences, trial differences, session differences and individual differences in EEG were all the different representations of energy pattern differences.
	
	(2) Due to the various degrees of differences, emotional differences and trial differences were hard to observe in multi-subjects scenes, which indicated both were the much smaller than session differences and individual differences.
	
	(3) The differences were ranked in ascending order: trial differences, emotional differences, session differences and individual differences.

	\subsection{The processing of different EEG information}
	The effects of different EEG information on emotion classification were different. Based on the above phenomena, we would like to eliminate the invalid information that hindered the work of the universal model and retained the beneficial information. The differences between individual and session domains caused by individual and session information are the major factors that hinder the generalization of the universal model. We applied the Personal-Zscore(PZ) in each session dataset to eliminate the individual differences and session differences in the dataset, which could unify different session domains into a domain with 0 mean and 1 variance to improve the model generalization ability\cite{chen2021personal}.
	
	Emotional information and trial information, which influence the sample distribution within the session domain, are the basis of emotion recognition. Based on previous work about the quantification of emotional information, it was found that different subjects' EEG data contained different levels of emotional information, but the results in Appendix \ref{appendixA} showed that even if data contained different degrees of emotional information and the impact of domain differences was eliminated, there were still unknown factors that led to large differences between different datasets. Therefore, both emotional information and trial information need more reasonable ways to improve the effectiveness of emotion recognition.

	\section{Weight-based Channel-model Matrix Framework}

	\subsection{Analyzing the distribution patterns of emotional difference features}
	After analyzing the causes of different EEG sample aggregation phenomena, we paid more attention to the challenge of cross-dataset emotion classification task and investigate the reasons for the poor performance of universal emotion recognition model. Specially, only the inverse power feature that reasonably represented energy was used for subsequent research.
	
	Firstly, it was necessary to quantify the emotional information of different dimensions, and we performed 2-class emotion intra-session classification with Leave-One-Trial-Out(LOTO) cross-validation for each dimensional dataset. Because only the inverse power feature datasets were selected, the dimension also could be considered as channel. The accuracies of each dimension were considered as weight, the demension(channel) with high weight was called as emotional difference feature. Therefore, weights were visualized in the same way as the energy pattern to observe the degree of emotional information for each channel and the distribution of emotional difference features.
	
	\begin{figure}[h]
		\centerline{\includegraphics[width=\linewidth]{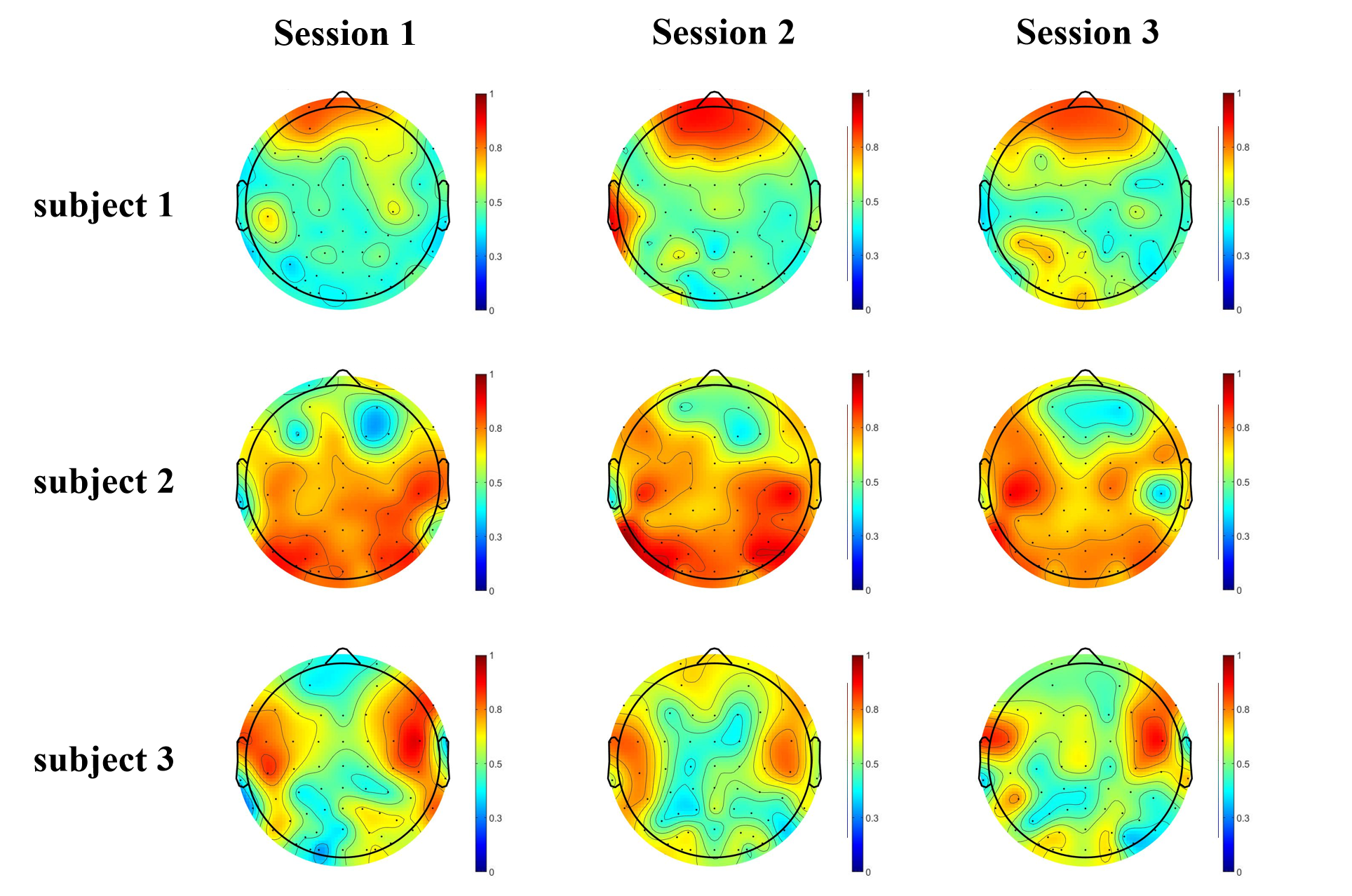}}
		\caption{Positive-negative emotional difference feature distribution of different subjects in different sessions.}
		\label{fig7}
	\end{figure}
	
	As shown in Fig. \ref{fig7}, there are significant differences between different subjects' emotional feature distributions. However, comparing the emotional feature distributions of sessions belonging to a single subject, these distribution patterns are similar. 
	
	Although the brain was a three-dimensional organic organization, the emotional feature distribution also could reflected the brain activities. we presented some assumptions from the perspective of the brain mechanism. For a single subject, the similar distributions between emotional features of sessions revealed that there existed neurons for emotional information processing in the brain. When brain were stimulated by the same emotion material in different time periods, the neurons produced some energy changes during the processing of emotional information, which were delivered to the similar scalp electrodes through stable brain circuits by EEG. For the difference of emotional feature distributions between subjects, we assumed that it would be caused by the complex individual factors, which should be validated through a further experiment.

	Next, we analyzed the impact of individual emotion feature distribution differences on the construction of a universal emotion recognition model from the classification perspective.
	
	(1) For the training set, it contains a mixture of emotional data and invalid data in each dimension. The emotional feature distribution differences cause the chaos of the sample distribution, it is difficult to learn the consistent emotional knowledge from the training set using machine learning algorithms. Consequently, the trained universal model have unstable emotion recognition performance.
	
	(2) For the test set, there exists a case that the test set would be misclassified by the traditional mutli-dimensional universal model, whose training data contains no or less data with similar emotional feature distributions. Therefore, the test set will obtain different results from the universal models constructed on different subjects' EEG data.  
	
	In summary, it was unreasonable to build the universal model with the data of different subjects without determining different subject emotional feature distributions, and the universal model obtained from such training process was also unreliable.
	
	Combining the results in Appendix A, we could know the impact of IEFDD on different datasets. Although all datasets had the high ratios of emotional features, the higher accuracy obtained with leave-one-session-out emotion classification in SEED was due to the high overlap of different subjects' emotional feature distributions. For the other datasets with low distribution overlap rate, IEFDD would extremely influence the classification accuracy.
	
	Therefore, there exists a large randomness for unknown subjects' emotion recognition without prior knowledge of the individual emotional feature distributions. When one test set was classified by models trained with different datasets separately, the classification results depended on the similarity of emotional feature distribution between training data and test data. This randomness is unacceptable for affective brain-computer interface. Therefore, we considered that the joint classification strategy of channel models might be better than the traditional multi-dimensional model classification in the EEG-based emotion recognition scene. Hence we proposed WCMF for dealing with the impact of individual emotional feature distribution difference, which would be a solution for cross-dataset emotion recognition.

	\subsection{Channel-model Matrix}
	Considering the individual emotional feature distribution difference, we employed the joint classification strategy of channel models. In this way, the training data of each channel model were selected based on the emotional feature distributions of different sessions. For each channel, the session data that showed emotional differences were retained, and invalid data were eliminated to ensure a stable emotional distribution of samples in the training set, and deep neural network(DNN) models with one hidden layer and 100 hidden-units were trained. In addition, the channel-model matrix was built up by the channel DNN model based on inverse power features, and this kind of modeling strategy could solve the problem of chaotic emotional sample distributions due to the individual emotional feature differences in traditional multidimensional modeling.
	
	However, there still exists the following limitations in classification with the channel-model matrix.
	
	(1) In terms of model training, it is necessary to design a method to quantify the emotional feature distribution of different subjects, which determines the quality of training data and the performances of channel models.
	
	(2) For emotion classification, although each channel model have the stable ability to recognize emotion, the test data are still composed of valid emotional information and unknown invalid information. Therefore, the classification results are influenced by the number of emotional difference features in the test set. The greater the number of emotional difference features, the greater number of channel model will judge the emotions correctly, and the greater positive contribution will be made for joint classification. On the contrary, invalid information would made greatly impacts to the joint classification, resulting in misclassification. 
	
	In summary, the classification of channel-model matrix still exists a randomness on the test set. Therefore, a efficient quantification method of emotional difference feature distribution would solve the randoness and improve the performance of channel-model matrix.
	
	\subsection{Exploration of weight extraction methods}
	For channel-model matrices, weights were beneficial for the selection of training data and the classification of known subjects, so it was necessary to find methods for learning weights from data in a reasonable and efficient way. Therefore, we proposed four methods for weight extraction: channel-based Classification(C) weight extraction method, DNN training-based DNN weight extraction method, inter-group difference-based T-test weight extraction method, and emotional sample relationship-based Correction T-test(CT) weight extraction method.
	\clearpage
	
	\subsubsection{Channel-based classificaion(C) weight extraction method}
	As mentioned in the Emotional Feature Distribution Pattern Analysis section, the single channel emotion classification task with LOTO cross-validation is performed on each session dataset, and the accuracies are considered as the C weight of each channel. However, due to the channel emotion classification task need to traverse each channel dataset, which is a very time consuming work.

	\subsubsection{DNN training-based DNN weight extraction method}
	Compared with the high cost of C weights, we refers to the attention mechanism for learning the emotional information of each dimension by DNN training, and generating the weight for each channel. The details of the DNN weight extraction method are as follows.
	
	\begin{algorithm}[htb]
	\caption{Calculation of the DNN weight}  
	\label{algthm2}
		\begin{algorithmic}[1]
			\Require
			$Data=\left \{emotion_{i}\right \}_{i=1}^{n}$, the emotion dataset with n classes. $Label$ the label of $Data$.
			\Ensure
			\State $pair=$\Call{powerSet}{$classes,2$}, produce the set of emotion pairs, $m$ pairs.
			\For{ index $j = 1 \to m$}  
			\State Initialize $DNNweight_{j}$ as zero vector.
			\State $data,label=$\Call{Divide}{$Data,Label,pair_{j}$}, get the j-th emotion pair data and label.
			\While {$accuracy \geq threshold$}
			\State $models,accuracy$=\Call{DNNTraining}{$data,label$,$k$}, train one hidden unint DNN model $k$ times.
			\State $hidden,weights$=\Call{DNNEncoder}{$models,data$}, get the hidden samples and weights for each DNN model.
			\State $W$, Select weights by the similar degree between $hidden$ and $label$.
			\State $weight$=\Call{fliter}{$W$,$wThreshold$}, retain the weight value $\ge$ $wThreshold$ and average weights. 
			\State $DNNweight_{j}=DNNweight_{j}+weight$, and add 1 to the positive items.
			\State $data$=\Call{reGroupData}{$data$}, retain the data with unextracted weights. 
			\EndWhile
			\EndFor
			\State \Return $DNNweight$
		\end{algorithmic} 
	\end{algorithm}  

	The idea of the DNN weight extraction method was to train a DNN model that only contained one hidden unit. 61-channel dataset were used to train the one-unit DNN, and this DNN would learn a way of information fusion from the dataset, where the edge weights between each channel data and the hidden unit described the contribution of each channel data to the hidden unit. Then, the higher weights were extracted, and the remaining channel data were used for the next round of weight extraction until training accuracy was lower than threshold.
	
	Specially, when performing DNN training multiple times in each round, there were two kind of opposite cases, positive and negative weight vectors. To eliminate this instability, based on the positive and negative weights, the hidden feature samples were obtained by positive and negative weight DNN model. Then, we compared the relationship between the positive and negative hidden feature sample distribution and selected the weight whose hidden feature sample relationship was similar to that of the label. In addition, the process of multi-times weight extraction still demanded a high time cost.
	
	\subsubsection{Inter-group difference-based T-test weight extraction method}
	To address the problem of higher computational costs, the t-test method was employed to measure the degree of inter-group differences between different emotions for each channel data. The algorithm is as follows.
	
	\begin{algorithm}[htb]
	\caption{Calculation of the T-test weight}  
	\label{algthm3}
		\begin{algorithmic}[1]
			\Require
			$Data=\left \{emotion_{i}\right \}_{i=1}^{n}$, the emotion dataset with n classes. $classes$ the set of emotion classes.
			\Ensure
			\State $pair=$\Call{powerSet}{$classes,2$}, produce the set of emotion pair, $m$ pair.
			\For{ index $j = 1 \to m$}  
			\State $p, q$=$pair_{j}$, get the index of j-th emotion pair.
			\State $p\_value$=\Call{ttest2}{$emotion_{p},emotion_{q}$}, perform t-test on two class dataset.
			\State $p\_value=\frac{1}{p\_value}$
			\State $weight_{j}$=\Call{LOGN}{$p\_value,10,2$}, apply $log_{10}$ function on $p\_value$ 2 times. 
			\EndFor
			\State \Return $weight$
		\end{algorithmic} 
	\end{algorithm}  
	
	Firstly, t-test was used to calculate the inter-group p-value of each channel data, and the smaller the p-value was, the greater degree of inter-group difference in that channel. However, because the value range of p-value was too small and the differences between the p-values of each channel were vast, which caused that directly applying the normalized p-value as weight could not measure the relationship of inter-group differences between channels effectively.
	
	Therefore, we performed a series of value range transformation operations, beginning with the transformation of the p-value into a positive integer value range by the inverse ratio operation. Facing the inverse p-values with the extremely large value range, we utilized the monotonicity and slope decreasing property of the log function to decrease the value range and retain the size relationship of p-values between each channel.

	Finally, the p-value after the value range transformation provided a more reasonable way to measure the degree of inter-group differences in each channel. In addition, the T-test weights had extremely low time costs.	
	
	\subsubsection{Comparison of C weight, DNN weight, T-test weight }
	In order to compare the performance of different weight extraction methods, the ideal cross-dataset classification task that labels were also used for weight extraction in training set and test set was performed. Because all datasets were balanced, the classification correct rate was selected as an evaluation metric for the three weight extraction methods. The results are as follows.
	
	\begin{table*}[t]
		\centering
		\caption{THE RESULT OF CROSS-DATASETS EMOTION RECOGNITION WITH THREE WEIGHT EXTRACTION METHODS}
		\label{tab3}
		\resizebox{!}{!}{%
			\begin{tabular}{ll|lll|lll|lll}
				\hline
				\multicolumn{2}{c|}{\textbf{Accuracy}} &
				\multicolumn{3}{c|}{\textbf{Classification weight}} &
				\multicolumn{3}{c|}{\textbf{DNN weight}} &
				\multicolumn{3}{c}{\textbf{T-test weight}} \\
				\textbf{Test} &
				\textbf{Train} &
				\textbf{Neg-Neu} &
				\textbf{Pos-Neu} &
				\textbf{Pos-Neg} &
				\textbf{Neg-Neu} &
				\textbf{Pos-Neu} &
				\textbf{Pos-Neg} &
				\textbf{Neg-Neu} &
				\textbf{Pos-Neu} &
				\textbf{Pos-Neg} \\ \hline
				\multirow{4}{*}{\textbf{SEED}}    & \textbf{SEED-IV} & 56.66\% & 79.65\% & 77.26\% & 61.64\% & 80.20\% & 76.96\% & 59.98\% & 79.09\% & 75.68\% \\
				& \textbf{SEED-V}  & 53.87\% & 80.43\% & 76.53\% & 61.53\% & 81.01\% & 76.83\% & 47.70\% & 78.26\% & 74.68\% \\
				& \textbf{RCLS}    & 58.17\% & 79.44\% & 77.63\% & 61.12\% & 81.13\% & 76.64\% & 56.55\% & 77.70\% & 75.33\% \\
				& \textbf{MPED}    & 56.97\% & 80.56\% & 77.14\% & 61.64\% & 81.39\% & 77.08\% & 56.05\% & 79.73\% & 76.94\% \\ \hline
				\multirow{4}{*}{\textbf{SEED-IV}} & \textbf{SEED}    & 54.20\% & 62.56\% & 59.91\% & 60.54\% & 67.59\% & 65.96\% & 53.76\% & 62.67\% & 59.39\% \\
				& \textbf{SEED-V}  & 46.26\% & 63.74\% & 60.15\% & 59.78\% & 67.50\% & 65.30\% & 46.72\% & 62.67\% & 59.63\% \\
				& \textbf{RCLS}    & 53.57\% & 63.70\% & 58.76\% & 59.46\% & 68.09\% & 65.65\% & 52.52\% & 62.93\% & 59.04\% \\
				& \textbf{MPED}    & 53.74\% & 63.37\% & 61.50\% & 59.98\% & 68.09\% & 66.02\% & 57.48\% & 63.65\% & 61.06\% \\ \hline
				\multirow{4}{*}{\textbf{SEED-V}}  & \textbf{SEED}    & 48.14\% & 64.32\% & 63.98\% & 54.46\% & 67.53\% & 68.82\% & 46.51\% & 64.41\% & 65.28\% \\
				& \textbf{SEED-IV} & 47.97\% & 63.80\% & 66.02\% & 53.91\% & 66.49\% & 69.03\% & 47.48\% & 63.96\% & 66.56\% \\
				& \textbf{RCLS}    & 51.01\% & 63.73\% & 66.74\% & 54.15\% & 67.66\% & 67.74\% & 47.90\% & 63.19\% & 67.81\% \\
				& \textbf{MPED}    & 48.44\% & 65.57\% & 66.15\% & 55.26\% & 68.06\% & 68.73\% & 50.80\% & 65.82\% & 68.61\% \\ \hline
				\multirow{4}{*}{\textbf{RCLS}}    & \textbf{SEED}    & 69.00\% & 67.36\% & 58.89\% & 70.32\% & 71.00\% & 63.50\% & 67.04\% & 67.18\% & 58.64\% \\
				& \textbf{SEED-IV} & 61.86\% & 69.46\% & 60.11\% & 69.39\% & 73.14\% & 63.61\% & 62.00\% & 69.57\% & 59.29\% \\
				& \textbf{SEED-V}  & 61.07\% & 69.18\% & 59.93\% & 70.46\% & 72.39\% & 64.46\% & 42.57\% & 67.00\% & 59.71\% \\
				& \textbf{MPED}    & 67.43\% & 71.61\% & 61.39\% & 67.68\% & 73.43\% & 64.14\% & 59.25\% & 70.36\% & 61.57\% \\ \hline
				\multirow{4}{*}{\textbf{MPED}}    & \textbf{SEED}    & 55.78\% & 62.13\% & 59.77\% & 60.07\% & 63.95\% & 63.07\% & 54.59\% & 61.28\% & 59.46\% \\
				& \textbf{SEED-IV} & 57.02\% & 62.70\% & 65.03\% & 59.33\% & 64.63\% & 64.40\% & 56.32\% & 62.46\% & 61.43\% \\
				& \textbf{SEED-V}  & 48.07\% & 64.36\% & 61.92\% & 61.75\% & 64.89\% & 62.00\% & 48.38\% & 61.89\% & 60.16\% \\
				& \textbf{RCLS}    & 54.89\% & 63.81\% & 63.41\% & 59.55\% & 64.66\% & 61.52\% & 54.73\% & 60.47\% & 61.82\% \\ \hline
				\multicolumn{2}{c|}{\textbf{Average}} &
				\textbf{55.21\%} &
				\textbf{68.07\%} &
				\textbf{65.11\%} &
				\textbf{61.10\%} &
				\textbf{70.64\%} &
				\textbf{67.57\%} &
				\textbf{53.42\%} &
				\textbf{67.21\%} &
				\textbf{64.60\%} \\ \hline
			\end{tabular}%
		}
	 	~\\
	 	~\\
	 	\leftline{\small \quad \quad \quad \quad Neg: negative emotion. Neu:neural emotion. Pos: Positive emotion.}
	\end{table*}

	As shown in Table \ref{tab3}, the results of T-test weight and C weight are similar, but both of them are lower than DNN weight. Comparing the similarities and differences of different weight extraction methods, we found that although the extracted weights could describe the differences between different emotion samples in each channel indeed, the similarities and differences of sample distribution tendencies on difference features were ignored. Taking the positive-negative emotion recognition scenario as an example, the following three sample relationships exist in dimensions.
	
	(1) Positive samples have higher feature values than negative samples.
	
	(2) Positive samples have lower feature values than negative samples.
	
	(3) There is no difference between positive samples and negative samples, and the samples are mixed up.
	
	For the channels reflecting such invalid sample relationships as (3), the chaos sample distributions are filtered out by the three weight extraction methods. However, the two types of sample relationships (1) and (2) are not determined on C weight and T-test weight extraction methods. Because these two relationships are mutually exclusive, it is reasonable for the model to classify the data with the same sample relationship as the training data, but the classification between data with different sample relationships would lead to great misclassification results.
	
	Concerning the DNN weight extraction method, the operation of selecting weights whose hidden feature samples have the same distribution with the label plays a role of sample relationship judgment. So, this is the reason why DNN weights outperformed than other weights, but the DNN weight extraction method still lacks the ability of processing the weights with the opposite relationship to the label.

	\subsubsection{Emotional sample relationship-based correction T-test(CT) weight extraction method}
	For the purpose of learning the undetermined emotional sample distribution information, we modified the T-test weight extraction method which had the lowest time cost, and the correction T-test weight extraction method based on the emotional sample relationship was proposed. The pseudo code is as follows.
	
	\begin{figure*}[t]
		\centerline{\includegraphics[width=\linewidth]{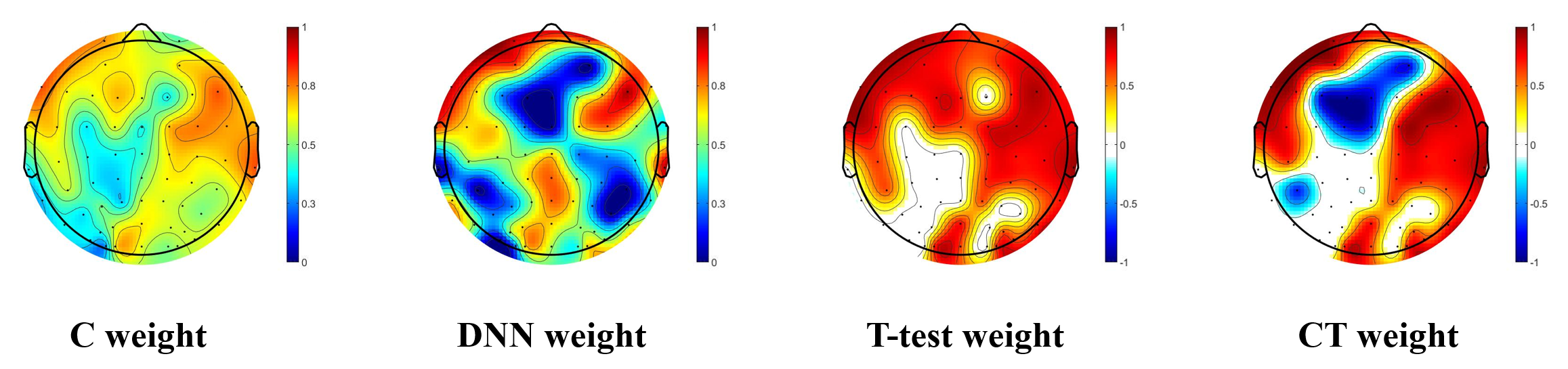}}
		\caption{Emotional difference feature distribution of different weight extraction methods. The value range of C, DNN and T-test weight are [0,1]. The value range of the CT weight is [-1,1], which represents the positive and negative sample relationships. }
		\label{fig8}
	\end{figure*}
	
	\begin{algorithm}[htb]
		\caption{Calculation of the correction T-test weight}  
		\label{algthm4}
		\begin{algorithmic}[1]
			\Require
			$Data=\left \{emotion_{i}\right \}_{i=1}^{n}$ the emotion dataset with n classes. $classes$ the set of emotion class. $Label$ the label of $Data$.
			\Ensure
			\State $Tweight,position$=\Call{TtestWeight}{$Data,classes$}, calculate the T-test weight and obtain the $m$ important channel positions.
			\For{ position index $j=1 \to m$}
			\For{ emotion index $k=1 \to n$}
			\State $sumFeature_{j,k}$=\Call{Mean}{$emotion_{k,j}$}, calculate the mean of the k-th emotion and j-th position feature values.
			\EndFor
			\If{$label$ size relationship consist to $sumFeature$}
			\State $coefficent_{j}$=1
			\Else
			\State $coefficent_{j}$=-1
			\EndIf
			\EndFor
			\State $weight$=$Tweight$ $\cdot$ $coefficient$
			\State \Return $weight$
		\end{algorithmic} 
	\end{algorithm}  
	
	The selection of emotional feature channels is performed by T-test weight exraction method. For each emotional feature channel, the averages of different emotions on inverse power feature are calculated, and the emotional averages are compared to define this channel’s feature emotional sample relationship. Positive sample relationship is the sample distribution consistent with label relationship, and negative in the contrary. We considered that the correction T-test weights could describe the individual emotional feature distribution pattern reasonably.

	As shown in Fig. \ref{fig8}, the higher weights are selected on all weights as the red part in all plots, but CT weight also selects high negative weight as the blue part in the plot of CT weight, and the white part in T-test and CT weight plots represents the channels with chaos information.
	
	
	\subsection{Comparison of modeling and weight extraction methods}
	To comprehensively analyze the impact of different modeling strategies and different weights on the classification results, we combined all classification results of different methods with the cross-dataset emotion classification task.
	
	\begin{table*}[b]
	\centering
	\caption{THE RESULTS OF CROSS-DARASET EMOTION RECOGNITION WITH DIFFERENT METHODS}
	\label{tab4}
	\resizebox{\textwidth}{!}{%
		\begin{tabular}{ll|cccccc|cccccccccccc}
			\hline
			\multicolumn{2}{c|}{\multirow{2}{*}{\textbf{Accuracy}}} &
			\multicolumn{6}{c|}{\textbf{Practical Scenarios}} &
			\multicolumn{12}{c}{\textbf{Weight-based Channel-model Matrix In Ideal Scenarios}} \\
			\multicolumn{2}{c|}{} &
			\multicolumn{3}{c}{\textbf{Classic Classification}} &
			\multicolumn{3}{c|}{\textbf{Matrix Classification}} &
			\multicolumn{3}{c}{\textbf{Classification weight}} &
			\multicolumn{3}{c}{\textbf{DNN weight}} &
			\multicolumn{3}{c}{\textbf{T-test weight}} &
			\multicolumn{3}{c}{\textbf{Correction T-test weight}} \\
			\textbf{Test} &
			\textbf{Train} &
			\multicolumn{1}{l}{\textbf{Neg-Neu}} &
			\multicolumn{1}{l}{\textbf{Pos-Neu}} &
			\multicolumn{1}{l}{\textbf{Pos-Neg}} &
			\multicolumn{1}{l}{\textbf{Neg-Neu}} &
			\multicolumn{1}{l}{\textbf{Pos-Neu}} &
			\multicolumn{1}{l|}{\textbf{Pos-Neg}} &
			\multicolumn{1}{l}{\textbf{Neg-Neu}} &
			\multicolumn{1}{l}{\textbf{Pos-Neu}} &
			\multicolumn{1}{l}{\textbf{Pos-Neg}} &
			\multicolumn{1}{l}{\textbf{Neg-Neu}} &
			\multicolumn{1}{l}{\textbf{Pos-Neu}} &
			\multicolumn{1}{l}{\textbf{Pos-Neg}} &
			\multicolumn{1}{l}{\textbf{Neg-Neu}} &
			\multicolumn{1}{l}{\textbf{Pos-Neu}} &
			\multicolumn{1}{l}{\textbf{Pos-Neg}} &
			\multicolumn{1}{l}{\textbf{Neg-Neu}} &
			\multicolumn{1}{l}{\textbf{Pos-Neu}} &
			\multicolumn{1}{l}{\textbf{Pos-Neg}} \\ \hline
			\multirow{4}{*}{\textbf{SEED}} &
			\textbf{SEED-IV} &
			53.80\% &
			74.40\% &
			\multicolumn{1}{c|}{67.76\%} &
			55.33\% &
			76.86\% &
			75.08\% &
			56.66\% &
			79.65\% &
			\multicolumn{1}{c|}{77.26\%} &
			61.64\% &
			80.20\% &
			\multicolumn{1}{c|}{76.96\%} &
			59.98\% &
			79.09\% &
			\multicolumn{1}{c|}{75.68\%} &
			73.76\% &
			82.81\% &
			82.44\% \\
			&
			\textbf{SEED-V} &
			51.48\% &
			64.14\% &
			\multicolumn{1}{c|}{75.97\%} &
			52.07\% &
			77.96\% &
			73.05\% &
			53.87\% &
			80.43\% &
			\multicolumn{1}{c|}{76.53\%} &
			61.53\% &
			81.01\% &
			\multicolumn{1}{c|}{76.83\%} &
			47.70\% &
			78.26\% &
			\multicolumn{1}{c|}{74.68\%} &
			74.22\% &
			83.39\% &
			81.16\% \\
			&
			\textbf{RCLS} &
			57.59\% &
			74.41\% &
			\multicolumn{1}{c|}{58.16\%} &
			56.35\% &
			77.03\% &
			75.09\% &
			58.17\% &
			79.44\% &
			\multicolumn{1}{c|}{77.63\%} &
			61.12\% &
			81.13\% &
			\multicolumn{1}{c|}{76.64\%} &
			56.55\% &
			77.70\% &
			\multicolumn{1}{c|}{75.33\%} &
			73.08\% &
			83.22\% &
			81.46\% \\
			&
			\textbf{MPED} &
			49.62\% &
			69.29\% &
			\multicolumn{1}{c|}{72.21\%} &
			55.51\% &
			78.20\% &
			74.51\% &
			56.97\% &
			80.56\% &
			\multicolumn{1}{c|}{77.14\%} &
			61.64\% &
			81.39\% &
			\multicolumn{1}{c|}{77.08\%} &
			56.05\% &
			79.73\% &
			\multicolumn{1}{c|}{76.94\%} &
			73.82\% &
			82.89\% &
			82.25\% \\ \hline
			\multirow{4}{*}{\textbf{SEED-IV}} &
			\textbf{SEED} &
			52.50\% &
			56.28\% &
			\multicolumn{1}{c|}{58.30\%} &
			54.15\% &
			60.54\% &
			57.80\% &
			54.20\% &
			62.56\% &
			\multicolumn{1}{c|}{59.91\%} &
			60.54\% &
			67.59\% &
			\multicolumn{1}{c|}{65.96\%} &
			53.76\% &
			62.67\% &
			\multicolumn{1}{c|}{59.39\%} &
			75.94\% &
			73.24\% &
			74.06\% \\
			&
			\textbf{SEED-V} &
			49.96\% &
			63.00\% &
			\multicolumn{1}{c|}{58.20\%} &
			46.35\% &
			61.63\% &
			58.17\% &
			46.26\% &
			63.74\% &
			\multicolumn{1}{c|}{60.15\%} &
			59.78\% &
			67.50\% &
			\multicolumn{1}{c|}{65.30\%} &
			46.72\% &
			62.67\% &
			\multicolumn{1}{c|}{59.63\%} &
			74.83\% &
			73.35\% &
			73.93\% \\
			&
			\textbf{RCLS} &
			55.94\% &
			63.98\% &
			\multicolumn{1}{c|}{57.13\%} &
			53.31\% &
			61.61\% &
			56.93\% &
			53.57\% &
			63.70\% &
			\multicolumn{1}{c|}{58.76\%} &
			59.46\% &
			68.09\% &
			\multicolumn{1}{c|}{65.65\%} &
			52.52\% &
			62.93\% &
			\multicolumn{1}{c|}{59.04\%} &
			74.61\% &
			73.93\% &
			74.13\% \\
			&
			\textbf{MPED} &
			54.00\% &
			61.30\% &
			\multicolumn{1}{c|}{58.28\%} &
			53.63\% &
			61.76\% &
			58.78\% &
			53.74\% &
			63.37\% &
			\multicolumn{1}{c|}{61.50\%} &
			59.98\% &
			68.09\% &
			\multicolumn{1}{c|}{66.02\%} &
			57.48\% &
			63.65\% &
			\multicolumn{1}{c|}{61.06\%} &
			75.11\% &
			73.91\% &
			74.06\% \\ \hline
			\multirow{4}{*}{\textbf{SEED-V}} &
			\textbf{SEED} &
			52.90\% &
			57.22\% &
			\multicolumn{1}{c|}{66.34\%} &
			47.76\% &
			61.08\% &
			62.45\% &
			48.14\% &
			64.32\% &
			\multicolumn{1}{c|}{63.98\%} &
			54.46\% &
			67.53\% &
			\multicolumn{1}{c|}{68.82\%} &
			46.51\% &
			64.41\% &
			\multicolumn{1}{c|}{65.28\%} &
			80.28\% &
			82.29\% &
			80.71\% \\
			&
			\textbf{SEED-IV} &
			51.08\% &
			61.75\% &
			\multicolumn{1}{c|}{59.74\%} &
			47.95\% &
			61.74\% &
			65.33\% &
			47.97\% &
			63.80\% &
			\multicolumn{1}{c|}{66.02\%} &
			53.91\% &
			66.49\% &
			\multicolumn{1}{c|}{69.03\%} &
			47.48\% &
			63.96\% &
			\multicolumn{1}{c|}{66.56\%} &
			80.59\% &
			83.91\% &
			83.84\% \\
			&
			\textbf{RCLS} &
			54.51\% &
			56.11\% &
			\multicolumn{1}{c|}{55.97\%} &
			49.34\% &
			61.74\% &
			65.92\% &
			51.01\% &
			63.73\% &
			\multicolumn{1}{c|}{66.74\%} &
			54.15\% &
			67.66\% &
			\multicolumn{1}{c|}{67.74\%} &
			47.90\% &
			63.19\% &
			\multicolumn{1}{c|}{67.81\%} &
			79.57\% &
			83.18\% &
			82.38\% \\
			&
			\textbf{MPED} &
			54.20\% &
			61.18\% &
			\multicolumn{1}{c|}{66.56\%} &
			48.00\% &
			62.67\% &
			65.57\% &
			48.44\% &
			65.57\% &
			\multicolumn{1}{c|}{66.15\%} &
			55.26\% &
			68.06\% &
			\multicolumn{1}{c|}{68.73\%} &
			50.80\% &
			65.82\% &
			\multicolumn{1}{c|}{68.61\%} &
			79.62\% &
			83.33\% &
			83.21\% \\ \hline
			\multirow{4}{*}{\textbf{RCLS}} &
			\textbf{SEED} &
			62.18\% &
			66.64\% &
			\multicolumn{1}{c|}{53.71\%} &
			64.57\% &
			64.57\% &
			58.00\% &
			69.00\% &
			67.36\% &
			\multicolumn{1}{c|}{58.89\%} &
			70.32\% &
			71.00\% &
			\multicolumn{1}{c|}{63.50\%} &
			67.04\% &
			67.18\% &
			\multicolumn{1}{c|}{58.64\%} &
			84.32\% &
			81.61\% &
			83.54\% \\
			&
			\textbf{SEED-IV} &
			56.14\% &
			62.07\% &
			\multicolumn{1}{c|}{63.18\%} &
			58.68\% &
			67.54\% &
			59.32\% &
			61.86\% &
			69.46\% &
			\multicolumn{1}{c|}{60.11\%} &
			69.39\% &
			73.14\% &
			\multicolumn{1}{c|}{63.61\%} &
			62.00\% &
			69.57\% &
			\multicolumn{1}{c|}{59.29\%} &
			85.14\% &
			83.71\% &
			86.25\% \\
			&
			\textbf{SEED-V} &
			51.04\% &
			57.86\% &
			\multicolumn{1}{c|}{60.75\%} &
			55.93\% &
			66.07\% &
			57.86\% &
			61.07\% &
			69.18\% &
			\multicolumn{1}{c|}{59.93\%} &
			70.46\% &
			72.39\% &
			\multicolumn{1}{c|}{64.46\%} &
			42.57\% &
			67.00\% &
			\multicolumn{1}{c|}{59.71\%} &
			83.82\% &
			82.61\% &
			84.71\% \\
			&
			\textbf{MPED} &
			52.86\% &
			63.75\% &
			\multicolumn{1}{c|}{60.68\%} &
			61.00\% &
			67.32\% &
			60.46\% &
			67.43\% &
			71.61\% &
			\multicolumn{1}{c|}{61.39\%} &
			67.68\% &
			73.43\% &
			\multicolumn{1}{c|}{64.14\%} &
			59.25\% &
			70.36\% &
			\multicolumn{1}{c|}{61.57\%} &
			85.18\% &
			84.04\% &
			85.43\% \\ \hline
			\multirow{4}{*}{\textbf{MPED}} &
			\textbf{SEED} &
			52.24\% &
			64.72\% &
			\multicolumn{1}{c|}{68.72\%} &
			54.94\% &
			59.30\% &
			57.77\% &
			55.78\% &
			62.13\% &
			\multicolumn{1}{c|}{59.77\%} &
			60.07\% &
			63.95\% &
			\multicolumn{1}{c|}{63.07\%} &
			54.59\% &
			61.28\% &
			\multicolumn{1}{c|}{59.46\%} &
			85.84\% &
			85.30\% &
			83.72\% \\
			&
			\textbf{SEED-IV} &
			56.82\% &
			64.83\% &
			\multicolumn{1}{c|}{65.40\%} &
			53.62\% &
			60.54\% &
			61.58\% &
			57.02\% &
			62.70\% &
			\multicolumn{1}{c|}{65.03\%} &
			59.33\% &
			64.63\% &
			\multicolumn{1}{c|}{64.40\%} &
			56.32\% &
			62.46\% &
			\multicolumn{1}{c|}{61.43\%} &
			85.67\% &
			87.43\% &
			85.03\% \\
			&
			\textbf{SEED-V} &
			55.45\% &
			62.60\% &
			\multicolumn{1}{c|}{66.42\%} &
			47.84\% &
			61.62\% &
			59.50\% &
			48.07\% &
			64.36\% &
			\multicolumn{1}{c|}{61.92\%} &
			61.75\% &
			64.89\% &
			\multicolumn{1}{c|}{62.00\%} &
			48.38\% &
			61.89\% &
			\multicolumn{1}{c|}{60.16\%} &
			86.58\% &
			86.53\% &
			85.00\% \\
			&
			\textbf{RCLS} &
			57.64\% &
			65.28\% &
			\multicolumn{1}{c|}{59.63\%} &
			54.39\% &
			61.01\% &
			59.36\% &
			54.89\% &
			63.81\% &
			\multicolumn{1}{c|}{63.41\%} &
			59.55\% &
			64.66\% &
			\multicolumn{1}{c|}{61.52\%} &
			54.73\% &
			60.47\% &
			\multicolumn{1}{c|}{61.82\%} &
			84.69\% &
			86.62\% &
			83.93\% \\ \hline
			\multicolumn{2}{c|}{\textbf{Average}} &
			\textbf{54.10\%} &
			\textbf{63.54\%} &
			\multicolumn{1}{c|}{\textbf{62.66\%}} &
			\textbf{53.54\%} &
			\textbf{65.54\%} &
			\textbf{63.13\%} &
			\textbf{55.21\%} &
			\textbf{68.07\%} &
			\multicolumn{1}{c|}{\textbf{65.11\%}} &
			\textbf{61.10\%} &
			\textbf{70.64\%} &
			\multicolumn{1}{c|}{\textbf{67.57\%}} &
			\textbf{53.42\%} &
			\textbf{67.21\%} &
			\multicolumn{1}{c|}{\textbf{64.60\%}} &
			\textbf{79.83\%} &
			\textbf{81.86\%} &
			\textbf{81.56\%} \\ \hline
		\end{tabular}%
	}
\end{table*}
	
	As shown in Table \ref{tab4}, we analyzed the results from different points of view. Comparing the classification results of different datasets, it could be found that the difficulties of cross-datasets classification tasks on different datasets were varied, which was caused by the individual emotional feature distribution difference. Among datasets, the SEED dataset obtained the best performance, which revealed that the subjects' emotional feature distribution had more common parts in SEED.

	\begin{figure}[h]
		\centering
		\includegraphics[width=\linewidth]{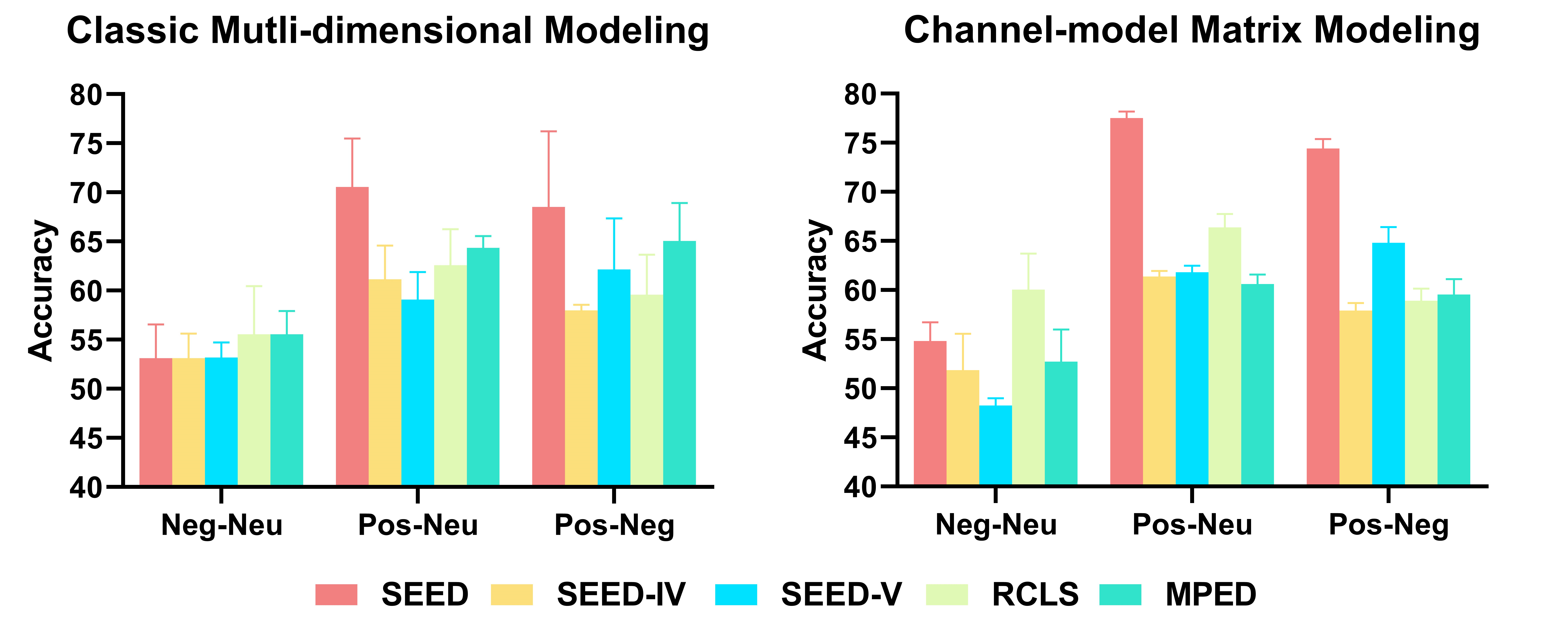}
		\caption{The results of different modeling methods on the cross-dataset emotion recognition task. }
		\label{fig9}
	\end{figure}
	
	\begin{figure*}[t]
		\centering
		\includegraphics[width=\linewidth]{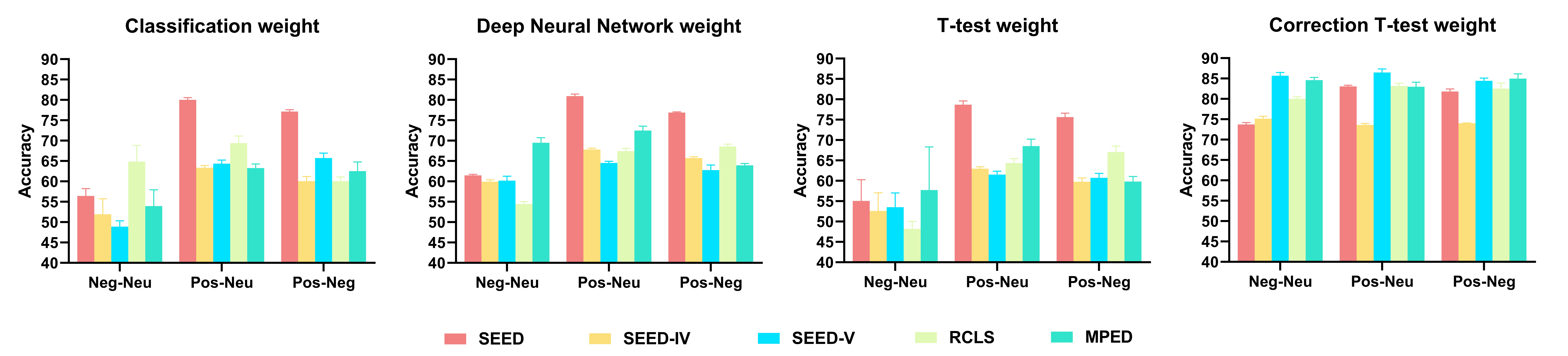}
		\caption{The results of different weight extraction methods on the cross-dataset emotion recognition task in ideal scenarios. }
		\label{fig10}
	\end{figure*}

	Comparing different modeling strategies, it could be found that the average classification results of channel-model matrix without the guidance of weights were slightly better than that of the multidimensional model (see Fig. \ref{fig9}). After observing the results of one test dataset under different training datasets, we found that the classification on the traditional multidimensional model was unstable, with higher variance, which was caused by the chaotic and complex sample space in the training set with IEFDD. In contrast, the classification accuracies were relatively stable for the channel-model matrix. The classification results are greatly influenced by the number of emotional features in test set, the more emotional features, the more number of channel models with correct classification, and the higher accuracy of joint classification.
	
	Comparing channel-model matrix classification and weight-based channel-model matrix classification, due to not all data had a great number of emotional difference features, the emotion recognition task for different subjects could not be handled by the channel-model matrix alone. Therefore, for the purpose of making the channel-model matrix work stably and effectively, the guidance of weight was necessary for the test data classification.
		
	Combining the classification results of different weights with the previous weights comparison analysis, it was found that the classification results of CT weights were the best(see Fig. \ref{fig10}), which validated the idea that sample relationship judgment in difference features was correct. Comparing the performance of different weight extraction methods, we considered that CT weight was the best extraction method, which not only had the T-test's extremely low computational cost but also had a more precise capability of learning emotional information than DNN weight.

	\subsection{Emotion state recognition using trial information}
	Considering that EEG had a stable trial sample aggregation phenomenon under power features, we considered trial information could be helpful for emotion sample recognition. Therefore, we tried to calibrate the classification results by real trial labels, and emotion sample recognition was turned into emotion state recognition. Firstly, true trial labels were used in cross-dataset classification with different weight extraction methods to define each trial sample set. Then, the predicted labels were unified to the majority of sample prediction for each trial sample set. 
	
	\begin{table}[h]
		\centering
		\caption{THE RESULTS OF DIFFERENT STRATEGY}
		\label{tab5}
		\resizebox{\linewidth}{!}{%
			\begin{tabular}{lcccccc}
				\hline
				\multicolumn{1}{c}{\multirow{3}{*}{\textbf{WEM}}} & \multicolumn{6}{c}{\textbf{Classification Strategy}}                                                             \\
				\multicolumn{1}{c}{}                              & \multicolumn{3}{c}{\textbf{Weight-based}}              & \multicolumn{3}{c}{\textbf{Weight \& trial correction}} \\
				\multicolumn{1}{c}{}                              & \textbf{Neg-Neu} & \textbf{Pos-Neu} & \textbf{Pos-Neg} & \textbf{Neg-Neu}  & \textbf{Pos-Neu} & \textbf{Pos-Neg} \\ \hline
				\textbf{C}  & 55.21\% & 68.07\% & 65.11\% & 61.10\% & 70.64\% & 67.57\% \\
				\textbf{DNN}    & 61.10\% & 70.64\% & 67.57\% & 64.33\% & 76.69\% & 72.33\% \\
				\textbf{T-test}  & 53.01\% & 68.28\% & 65.61\% & 54.61\% & 74.12\% & 69.86\% \\
				\textbf{CT-test} & 79.83\% & 81.86\% & 81.56\% & 86.88\% & 88.79\% & 88.62\% \\ \hline
			\end{tabular}%
		
		}
		~\\
	    \leftline {\small \quad \quad WEM: weight extraction method. C: classification weight. }
	    \leftline {\small DNN: deep neural network weight.}
	   
	\end{table}
	
	As shown in Table \ref{tab5}, two points could be observed.
	
	(1) With the help of trial information, the emotion state classification got an improvement compared with the emotion sample classification. This phenomenon indicated that there were more or fewer misclassifications in trial datasets.
	
	(2) The improvement brought by the trial correction strategy depended on how great the emotion sample classification results were. When the majority of samples in the trial clusters were classified correctly, the misclassified samples could be corrected. Therefore, this strategy usually brought a boost to the case with high classification correct rate on the emotion sample recognition.
	
	To solve the situation that true trial labels were not available, we applied the DBSCAN algorithm for the estimation of trial number in unknown datasets based on the trial aggregation of power features. Then, the predicted trial labels were obtained by k-means algorithm. The details of trial label estimation algorithm are shown in Appendix C.
	 
	~\\
	
	\section{Experiments and results}
	As mentioned in the above sections, all the weight extraction methods are based on labels, so weights extracted from the training set are reasonable. However, in practical scenarios, the labels are unknown for EEG generated in real time, and the weight extraction methods can not be employed directly. Moreover, the cross-dataset emotion classification task is still tricky and the practical scenarios of aBCI are variable. Here, we validated the performance of WCMF for the cross-dataset task in two kinds of practical scenarios.
	
	\subsection{Emotion recognition in the unknown subject scenario}
	In this scene, the emotion label and trial label are unknown for the test data. However, since weight extraction method required labels, we predicted the labels based on the emotional aggregation phenomenon and trial aggregation phenomenon under the inverse power feature.    
	
	Firstly, the emotion cluster label was generated by the combination of t-sne and k-means methods. The result of clustering depended on the ratio of emotion features and the amount of emotional information in the data, which would affect the accuracy of k-means algorithm for clustering emotion samples. Because the k-means algorithm only served the function of clustering, the relationship between the labels could not be determined. For this reason, we extracted DNN weights based on two kinds of possible labels, and the T-test weights were calculated on the basis of emotion cluster labels. Comparing the similarity between the two DNN weights and the T-test weights, the labels of DNN weights with high similarity to T-test weights were considered as the predicted labels for CT weight extraction. Secondly, predicted trial labels were estimated by the trial label estimation algorithm, which was mentioned in Appendix C. Finally, the predicted CT weights and predicted trial labels were combined to classify the unknown subjects' emotions. The above experiment were based on cross-dataset 2-class emotion classification tasks. The results are shown in Table \ref{tab6}.
	
	\begin{table*}[t]
		\centering
		\caption{THE RESULTS OF EMOTION RECOGNITION IN PRACTICAL SCENE}
		\label{tab6}
		\resizebox{!}{!}{%
			\begin{tabular}{ll|lll|lll}
				\hline
				\multicolumn{2}{c|}{\textbf{Accuracy}} & \multicolumn{3}{c|}{\textbf{Matrix classification}}    & \multicolumn{3}{c}{\textbf{Weight \& trial correction}} \\
				\textbf{Test}     & \textbf{Train}     & \textbf{Neg-Neu} & \textbf{Pos-Neu} & \textbf{Pos-Neg} & \textbf{Neg-Neu}  & \textbf{Pos-Neu} & \textbf{Pos-Neg} \\ \hline
				\multirow{4}{*}{\textbf{SEED}}    & \textbf{SEED-IV} & 54.08\% & 73.05\% & 73.22\% & 60.18\% & 80.22\% & 80.64\% \\
				& \textbf{SEED-V}  & 45.83\% & 76.35\% & 72.45\% & 60.51\% & 82.94\% & 80.64\% \\
				& \textbf{RCLS}    & 54.85\% & 75.60\% & 73.53\% & 59.93\% & 82.12\% & 80.76\% \\
				& \textbf{MPED}    & 52.33\% & 70.29\% & 70.93\% & 60.45\% & 80.42\% & 80.80\% \\ \hline
				\multirow{4}{*}{\textbf{SEED-IV}} & \textbf{SEED}    & 52.93\% & 60.52\% & 57.61\% & 52.78\% & 59.70\% & 56.81\% \\
				& \textbf{SEED-V}  & 46.69\% & 61.11\% & 57.65\% & 52.46\% & 59.72\% & 55.83\% \\
				& \textbf{RCLS}    & 52.07\% & 60.94\% & 57.15\% & 53.56\% & 61.70\% & 56.13\% \\
				& \textbf{MPED}    & 50.35\% & 58.11\% & 56.43\% & 52.26\% & 61.00\% & 56.11\% \\ \hline
				\multirow{4}{*}{\textbf{SEED-V}}  & \textbf{SEED}    & 45.49\% & 59.95\% & 61.68\% & 48.19\% & 58.00\% & 60.45\% \\
				& \textbf{SEED-IV} & 46.79\% & 59.17\% & 63.68\% & 46.74\% & 55.28\% & 60.19\% \\
				& \textbf{RCLS}    & 48.63\% & 60.90\% & 65.03\% & 47.95\% & 57.15\% & 60.17\% \\
				& \textbf{MPED}    & 48.89\% & 57.90\% & 58.45\% & 47.53\% & 56.56\% & 60.50\% \\ \hline
				\multirow{4}{*}{\textbf{RCLS}}    & \textbf{SEED}    & 56.14\% & 63.64\% & 56.96\% & 69.89\% & 75.07\% & 61.25\% \\
				& \textbf{SEED-IV} & 57.86\% & 66.04\% & 58.14\% & 69.89\% & 77.64\% & 59.79\% \\
				& \textbf{SEED-V}  & 38.93\% & 64.79\% & 57.86\% & 68.50\% & 78.46\% & 62.32\% \\
				& \textbf{MPED}    & 55.54\% & 62.93\% & 55.68\% & 69.18\% & 76.25\% & 59.86\% \\ \hline
				\multirow{4}{*}{\textbf{MPED}}    & \textbf{SEED}    & 54.87\% & 59.09\% & 57.06\% & 54.52\% & 68.37\% & 61.12\% \\
				& \textbf{SEED-IV} & 53.59\% & 57.68\% & 59.12\% & 55.57\% & 63.91\% & 60.10\% \\
				& \textbf{SEED-V}  & 47.16\% & 59.69\% & 56.53\% & 56.11\% & 67.91\% & 60.97\% \\
				& \textbf{RCLS}    & 53.65\% & 59.23\% & 58.11\% & 55.07\% & 66.99\% & 61.04\% \\ \hline
				\multicolumn{2}{c|}{\textbf{Average}}  & \textbf{50.83\%} & \textbf{63.35\%} & \textbf{61.36\%} & \textbf{57.06\%}  & \textbf{68.47\%} & \textbf{63.77\%} \\ \hline
			\end{tabular}%
		}
	\end{table*}
	
	Comparing the results in Table \ref{tab6}, we found that the predicted trial labels and CT weights had different impacts for different datasets, there were different degrees of improvements in performance for SEED, RCLS, and MPED datasets. The cases of accuracy reduction occurred on SEED-IV and SEED-V datasets.
	
	Combining the results of weight-based cross-dataset experiments in ideal scenarios, it is known that the classification of unknown subjects greatly depended on the effectiveness of weights. The similarity between the predicted weights and the true weights influences the judgement of joint emotion classification directly.
	
	For each session dataset, the performance of emotion label estimation determines the results of WCMF indirectly. When there exists strong emotional differences in the session dataset, the clustering algorithm is able to perform emotion clustering well, and the reliable emotion estimation labels were obtained. In this way, the knowledge of weights based on reliable predicted labels is closer to the knowledge learned in the ideal scenarios, which improves the emotion recognition ability on unknown subjects. In contrast, for the session dataset with poor emotional information, it is difficult to perform emotion clustering in the sample space with chaotic emotion distributions, and each emotion cluster obtained by k-means contains a certain number of other emotion samples. Therefore, weights based on chaotic predicted labels are unable to represent the differences between different emotions correctly. Regarding the results of SEED-IV and SEED-V as an example, due to the low ratio of emotional difference features, they got greater impact by invalid feature, which caused the chaos of sample space, the mistake of label estimation and invalid weight extraction. Finally, the invalid weight played a negative role in the cross-dataset emotion recognition.
	
	In summary, the results reveals that emotion recognition for completely unknown subjects is very tricky due to IEFDD. Although the strategy of label predicteding makes some effect, it still exists some limitations, which extremely relies on the emotional information of data.
	~\\
	\subsection{Emotion recognition based on the prior weights scenario}
	Considering the challenge of emotion recognition in unknown subject scenarios, it was unreliable to recognize emotion just relying on data. Therefore, based on the characteristic of the similar session emotional feature distributions belonging to one subject, the emotion recognition based on the prior weights was performed. The visualizations of CT weights for different subjects' sessions are as follows.
	
	\begin{figure}[htb]
		\centering
		\includegraphics[width=\linewidth]{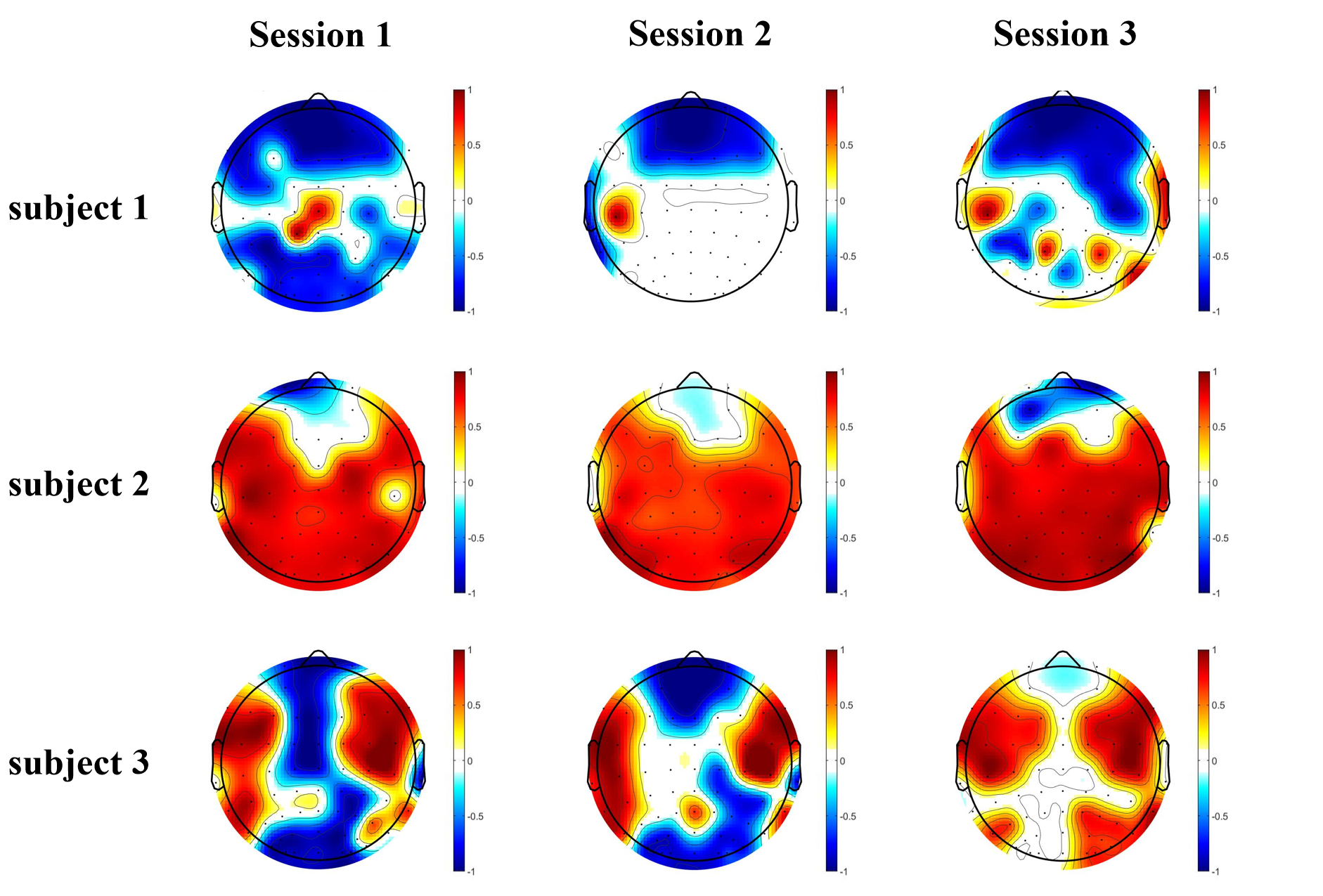}
		\caption{Comparison of CT weights between different subjects and different sessions.}
		\label{fig11}
	\end{figure}
	
	\begin{table*}[b]
		\centering
		\caption{THE RESULTS OF CLASSIFICATION WITH PRIOR WEIGHT}
		\label{tab7}
		\begin{tabular}{clllllll}
			\hline
			\multicolumn{2}{c}{\textbf{Accuracy}} &
			\multicolumn{3}{c}{\textbf{Trial correction}} &
			\multicolumn{3}{c}{\textbf{\begin{tabular}[c]{@{}c@{}}Weight \& trial correction\end{tabular}}} \\
			\multicolumn{1}{l}{\textbf{Test}} &
			\textbf{Train} &
			\textbf{Neg-Neu} &
			\textbf{Pos-Neu} &
			\textbf{Pos-Neg} &
			\textbf{Neg-Neu} &
			\textbf{Pos-Neu} &
			\textbf{Pos-Neg} \\ \hline
			\multirow{4}{*}{\textbf{SEED}}    & \textbf{SEED-IV} & 55.56\% & 78.67\% & 79.33\% & 64.67\% & 85.11\% & 86.44\% \\
			& \textbf{SEED-V}  & 42.00\% & 84.44\% & 78.00\% & 60.22\% & 89.33\% & 86.00\% \\
			& \textbf{RCLS}    & 58.67\% & 84.00\% & 80.00\% & 64.89\% & 86.44\% & 85.11\% \\
			& \textbf{MPED}    & 50.22\% & 75.33\% & 77.11\% & 63.78\% & 86.67\% & 85.78\% \\ \hline
			\multirow{4}{*}{\textbf{SEED-IV}} & \textbf{SEED}    & 52.04\% & 64.63\% & 60.00\% & 66.11\% & 64.44\% & 69.07\% \\
			& \textbf{SEED-V}  & 44.26\% & 63.33\% & 58.52\% & 65.56\% & 62.78\% & 67.59\% \\
			& \textbf{RCLS}    & 51.48\% & 64.07\% & 57.59\% & 65.56\% & 63.52\% & 67.96\% \\
			& \textbf{MPED}    & 50.93\% & 56.48\% & 58.15\% & 66.67\% & 62.04\% & 66.67\% \\ \hline
			\multirow{4}{*}{\textbf{SEED-V}}  & \textbf{SEED}    & 47.92\% & 63.89\% & 63.54\% & 58.33\% & 71.53\% & 71.88\% \\
			& \textbf{SEED-IV} & 45.49\% & 63.54\% & 69.79\% & 59.03\% & 70.14\% & 67.01\% \\
			& \textbf{RCLS}    & 46.88\% & 66.67\% & 69.44\% & 60.76\% & 71.18\% & 71.18\% \\
			& \textbf{MPED}    & 48.96\% & 57.99\% & 60.07\% & 59.72\% & 69.44\% & 70.83\% \\ \hline
			\multicolumn{2}{c}{\textbf{Average}}                 & \textbf{49.53\%} & \textbf{68.59\%} & \textbf{67.63\%} & \textbf{62.94\%} & \textbf{73.55\%} & \textbf{74.63\%} \\ \hline
		\end{tabular}%
	\end{table*}
	
	As shown in Fig. \ref{fig11}, the visualization of CT weights had similar results as the previous visualization of C weight, where the different subject’s weights existed large differences, and the session CT weights conducted from a single subject showed similar emotional feature distribution patterns. It was known that CT weights measured the brain activity under different emotion stimulation. However, due to the complexity of neural activity, EEG not only responded to main brain activity evoked by emotion, but also contained other unknown brain activity, which led great similarities and some differences between different session weights from a single subject. 
	
	To capture stable emotional feature distribution pattern, it is necessary to perform CT weight extraction from multiple times emotion recognition tasks, and the common parts between CT weights are maintained as a stable and valid emotional feature distribution pattern for that subject. Thus, the affective brain-computer interface would work as a system based on initial weights, which are used to analyze the data acquired through online emotion recognition tasks. After each work session, initial weights would be more stable through the updating based on the online session data constantly. 
	
	For the existing data, we picked the datasets with 3 sessions to validate the effect of weight merging strategy, and leave-one-session-out strategy was employed for the generation of unknown session's prior weight. For a subject dataset, when one session is considered as test data, the other sessions are considered as priori data, and the CT weight extraction method is employed on each priori session relatively. Then, the common parts between priori CT weights are merged as test session’s weight, which simulates a process of initial weight updating. Because the subject weights are generated from the prior knowledge, the problem of information leakage would not happen. Finally, we took SEED, SEED-IV, and SEED-V as the test sets, and performed cross-dataset 2-class emotion state classification task based on the prior weights.
	
	As shown in Table \ref{tab7}, due to the similarity of CT weights for different sessions of the same subject, the prior weights served as guidance for the emotion recognition task of unknown sessions. In general, the average accuracy of the SEED, SEED-IV, and SEED-V datasets got an improvement of 13.41\%, 4.96\%, and 7.00\% on negative-neutral, positive-neutral, and positive-negative under the guidance of prior weights.

    \section{Discussion}
	Comparing the cross-dataset emotion classification results under two kinds of practical application scenarios, it was found that emotion recognition of priori weights had higher reliability than that of predicteding weights. Considering the challenges of EEG-based cross-dataset emotion recognition, we believes that the strategy based on prior weights and updating weights continuously is a reasonable solution for affective brain-computer interfaces in practical scenarios. In the offline phase, the initial weights are extracted through one or more emotion experiments for unknown subjects. In the online phase, the weights will guide the model matrix to perform stable emotion recognition in real time. Beside, the EEG data obtained from the online working session are used for the updating of model matrix and weights, which not only expands the data of channel model, but also makes the subjects’ weight better. As the affective brain-computer interface works over and over again, an extremely stable model matrix will be obtained, and different subjects’ effective emotional feature distribution pattern will provide a reference for further research about gender\cite{zhu2015cross}, behavior\cite{waschke2021behavior} and so on.
	
	Compared with the traditional multidimensional universal model, the following advantages of WCMF are found.
	
	(1) The joint classification of channel-model matrix avoids the risk of overfitting and the curse of dimensionality, and the classification principle of channel-model matrix is more reasonable, the performances are more stable under the guidance of the weights.
	
	(2) In terms of universal model training, CT weights not only learn the emotional feature distribution for different subjects reasonably and accurately, but also have lower time cost and computational cost. Under the guidance of weights, a certain amount of data with consistent emotional sample distribution is selected for the training of channel models, which not only accelerates the models' training, but also makes each channel model have stable emotional classification capability. 
	
	(3) In practical application scenarios, EEG data generated in real time can be recognized by different model matrices with the help of weights, even though the model matrices built by different groups of subjects. Importantly, the storage of weights is more convenient than the storage of a priori EEG signals or EEG feature data sets.
	
	Although WCMF has some advantages, some limitations of WCMF and EEG-based affective computing are found through these mutli-dataset works in this article. 
	
	For the construction of datasets, due to the cross-time energy pattern during trial and the trial sample aggregation phenomenon, we consider that the number of trial is more important than the length of trial in this work. Therefore, we suggest that the construction of dataset should be goal-oriented. For the purpose of multi emotion analysis, we suggest to expand the number of trial with a certain length, which could make the result more stable and conclusions of emotional differences more reliable. In addition, for the purpose of further analysis on EEG individual difference, the dataset needs more number of subject and session. For the purpose of real-time emotion degree change analysis, the length of trial and the reasonable acquisition method of real-time emotion changing label are more important.	
	
	For multi-emotion recognition, we investigated the differences between different emotion pairs in this article, although CT weight extraction method had the ability to capture differences between different emotion pair, and whether there exist stable patterns of multi-emotion relationships in different subjects still need more data and experiments for further investigation. Combining the phenomenon of few differences between negative and neutral emotions, it’s necessary to validate that this phenomenon is caused by the incompletely induction of negative emotion by the emotional stimulus material, or due to the brain activity does not have significant differences between negative and neutral emotions\cite{li2018hierarchical}. Therefore, we believe that emotion recognition in multi-emotion scenarios is more challenging than cross-dataset tasks, not only needs to find multi-emotion relationships among different subjects, but also needs to verify the reasonableness of multi-emotion recognition.
	
	For application of affective brain-computer interface, it’s unstable, unreliable and impossible for multi-dimensional universal model without weight guidance to deal all emotion recognition task in various application scene of emotion recognition. Although the CT-test weight-based channel-model matrix framework can handle the tricky cross-dataset emotion recognition task, this solution still exists so many limitations that framework need more improvements to fit the various affective computing application scenes. After all, there are not all scenarios need to process all emotions in real life. Considering the performance of the universal model, the challenge of emotion recognition and the complexity of application scenarios, it’s necessary to specify different application scenes. Based on the emotion recognition requirements of different application scenarios, the potential emotions in different scenarios will be analyzed. Then, customized emotion subset modeling solutions will be designed to build a stable and effective emotion subset model for emotional sub-scenes. Of course, the multi-emotion universal model is still the final goal of affective brain-computer interface, but the sub-scene emotion recognition modeling can provide valuable experiences for EEG-based affective computing.
	
	
	\section{Acknowledgments}
	This work was supported in part by the National Key Research and Development Program of China [Grant No.2019YFA0706200], in part by Scientific and technological innovation 2030 project of MOST [No.2021ZD0202000, No.2022ZD0208500, No.2021ZD0200601], in part by the National Natural Science Foundation of China [Grant No.62102172, No.61627808, No.61632014], in part by  Gansu Province Science and Technology Program [No.22JR5RA489], in part by Key Program of Natural Science Foundation of Gansu Province [No.22JR5RA410], in part by the Typical Application Demon-stration Project of Shandong Academy of Intelligent Computing Technology [No.SDAICT2081020],  and the Fundamental Research Funds for the Central Universities [No.lzujbky-2022-10, No.lzujbky-2018-125].
	
	\bibliographystyle{IEEEtran}
	\bibliography{ref}

\begin{thebibliography}{10}
\providecommand{\url}[1]{#1}
\csname url@samestyle\endcsname
\providecommand{\newblock}{\relax}
\providecommand{\bibinfo}[2]{#2}
\providecommand{\BIBentrySTDinterwordspacing}{\spaceskip=0pt\relax}
\providecommand{\BIBentryALTinterwordstretchfactor}{4}
\providecommand{\BIBentryALTinterwordspacing}{\spaceskip=\fontdimen2\font plus
\BIBentryALTinterwordstretchfactor\fontdimen3\font minus
  \fontdimen4\font\relax}
\providecommand{\BIBforeignlanguage}[2]{{%
\expandafter\ifx\csname l@#1\endcsname\relax
\typeout{** WARNING: IEEEtran.bst: No hyphenation pattern has been}%
\typeout{** loaded for the language `#1'. Using the pattern for}%
\typeout{** the default language instead.}%
\else
\language=\csname l@#1\endcsname
\fi
#2}}
\providecommand{\BIBdecl}{\relax}
\BIBdecl

\bibitem{huang2019eeg}
H.~Huang, Q.~Xie, J.~Pan, Y.~He, Z.~Wen, R.~Yu, and Y.~Li, ``An eeg-based brain
  computer interface for emotion recognition and its application in patients
  with disorder of consciousness,'' \emph{IEEE Transactions on Affective
  Computing}, vol.~12, no.~4, pp. 832--842, 2019.

\bibitem{wei2020eeg}
C.~Wei, L.-l. Chen, Z.-z. Song, X.-g. Lou, and D.-d. Li, ``Eeg-based emotion
  recognition using simple recurrent units network and ensemble learning,''
  \emph{Biomedical Signal Processing and Control}, vol.~58, p. 101756, 2020.

\bibitem{peng2021self}
Y.~Peng, W.~Kong, F.~Qin, F.~Nie, J.~Fang, B.-L. Lu, and A.~Cichocki,
  ``Self-weighted semi-supervised classification for joint eeg-based emotion
  recognition and affective activation patterns mining,'' \emph{IEEE
  Transactions on Instrumentation and Measurement}, vol.~70, pp. 1--11, 2021.

\bibitem{chen2021ms}
H.~Chen, M.~Jin, Z.~Li, C.~Fan, J.~Li, and H.~He, ``Ms-mda: Multisource
  marginal distribution adaptation for cross-subject and cross-session eeg
  emotion recognition,'' \emph{Frontiers in Neuroscience}, vol.~15, 2021.

\bibitem{fdez2021cross}
J.~Fdez, N.~Guttenberg, O.~Witkowski, and A.~Pasquali, ``Cross-subject
  eeg-based emotion recognition through neural networks with stratified
  normalization,'' \emph{Frontiers in neuroscience}, vol.~15, p. 626277, 2021.

\bibitem{shen2022contrastive}
X.~Shen, X.~Liu, X.~Hu, D.~Zhang, and S.~Song, ``Contrastive learning of
  subject-invariant eeg representations for cross-subject emotion
  recognition,'' \emph{IEEE Transactions on Affective Computing}, 2022.

\bibitem{huang2022generator}
D.~Huang, S.~Zhou, and D.~Jiang, ``Generator-based domain adaptation method
  with knowledge free for cross-subject eeg emotion recognition,''
  \emph{Cognitive Computation}, pp. 1--12, 2022.

\bibitem{cimtay2020investigating}
Y.~Cimtay and E.~Ekmekcioglu, ``Investigating the use of pretrained
  convolutional neural network on cross-subject and cross-dataset eeg emotion
  recognition,'' \emph{Sensors}, vol.~20, no.~7, p. 2034, 2020.

\bibitem{lan2018domain}
Z.~Lan, O.~Sourina, L.~Wang, R.~Scherer, and G.~R. M{\"u}ller-Putz, ``Domain
  adaptation techniques for eeg-based emotion recognition: a comparative study
  on two public datasets,'' \emph{IEEE Transactions on Cognitive and
  Developmental Systems}, vol.~11, no.~1, pp. 85--94, 2018.

\bibitem{he2022adversarial}
Z.~He, Y.~Zhong, and J.~Pan, ``An adversarial discriminative temporal
  convolutional network for eeg-based cross-domain emotion recognition,''
  \emph{Computers in biology and medicine}, vol. 141, p. 105048, 2022.

\bibitem{yin2021eeg}
Y.~Yin, X.~Zheng, B.~Hu, Y.~Zhang, and X.~Cui, ``Eeg emotion recognition using
  fusion model of graph convolutional neural networks and lstm,'' \emph{Applied
  Soft Computing}, vol. 100, p. 106954, 2021.

\bibitem{huang2021differences}
D.~Huang, S.~Chen, C.~Liu, L.~Zheng, Z.~Tian, and D.~Jiang, ``Differences first
  in asymmetric brain: A bi-hemisphere discrepancy convolutional neural network
  for eeg emotion recognition,'' \emph{Neurocomputing}, vol. 448, pp. 140--151,
  2021.

\bibitem{he2022cross}
Z.~He, N.~Zhuang, G.~Bao, Y.~Zeng, and B.~Yan, ``Cross-day eeg-based emotion
  recognition using transfer component analysis,'' \emph{Electronics}, vol.~11,
  no.~4, p. 651, 2022.

\bibitem{lin2019constructing}
Y.-P. Lin, ``Constructing a personalized cross-day eeg-based
  emotion-classification model using transfer learning,'' \emph{IEEE journal of
  biomedical and health informatics}, vol.~24, no.~5, pp. 1255--1264, 2019.

\bibitem{bao2021two}
G.~Bao, N.~Zhuang, L.~Tong, B.~Yan, J.~Shu, L.~Wang, Y.~Zeng, and Z.~Shen,
  ``Two-level domain adaptation neural network for eeg-based emotion
  recognition,'' \emph{Frontiers in Human Neuroscience}, vol.~14, p. 605246,
  2021.

\bibitem{cai2022cross}
Q.~Cai, J.-P. An, H.-Y. Li, J.-Y. Guo, and Z.-K. Gao, ``Cross-subject emotion
  recognition using visibility graph and genetic algorithm-based convolution
  neural network,'' \emph{Chaos: An Interdisciplinary Journal of Nonlinear
  Science}, vol.~32, no.~9, p. 093110, 2022.

\bibitem{meng2022deep}
M.~Meng, J.~Hu, Y.~Gao, W.~Kong, and Z.~Luo, ``A deep subdomain associate
  adaptation network for cross-session and cross-subject eeg emotion
  recognition,'' \emph{Biomedical Signal Processing and Control}, vol.~78, p.
  103873, 2022.

\bibitem{zhao2021plug}
L.-M. Zhao, X.~Yan, and B.-L. Lu, ``Plug-and-play domain adaptation for
  cross-subject eeg-based emotion recognition,'' in \emph{Proceedings of the
  AAAI Conference on Artificial Intelligence}, vol.~35, no.~1, 2021, pp.
  863--870.

\bibitem{luo2018wgan}
Y.~Luo, S.-Y. Zhang, W.-L. Zheng, and B.-L. Lu, ``Wgan domain adaptation for
  eeg-based emotion recognition,'' in \emph{International Conference on Neural
  Information Processing}.\hskip 1em plus 0.5em minus 0.4em\relax Springer,
  2018, pp. 275--286.

\bibitem{li2022cross}
J.~Li, H.~Hua, Z.~Xu, L.~Shu, X.~Xu, F.~Kuang, and S.~Wu, ``Cross-subject eeg
  emotion recognition combined with connectivity features and meta-transfer
  learning,'' \emph{Computers in Biology and Medicine}, vol. 145, p. 105519,
  2022.

\bibitem{wang2021deep}
F.~Wang, W.~Zhang, Z.~Xu, J.~Ping, and H.~Chu, ``A deep multi-source adaptation
  transfer network for cross-subject electroencephalogram emotion
  recognition,'' \emph{Neural Computing and Applications}, vol.~33, no.~15, pp.
  9061--9073, 2021.

\bibitem{li2021cross}
J.~Li, S.~Li, J.~Pan, and F.~Wang, ``Cross-subject eeg emotion recognition with
  self-organized graph neural network,'' \emph{Frontiers in Neuroscience}, p.
  689, 2021.

\bibitem{zheng2015investigating}
W.-L. Zheng and B.-L. Lu, ``Investigating critical frequency bands and channels
  for eeg-based emotion recognition with deep neural networks,'' \emph{IEEE
  Transactions on autonomous mental development}, vol.~7, no.~3, pp. 162--175,
  2015.

\bibitem{duan2013differential}
R.-N. Duan, J.-Y. Zhu, and B.-L. Lu, ``Differential entropy feature for
  eeg-based emotion classification,'' in \emph{2013 6th International IEEE/EMBS
  Conference on Neural Engineering (NER)}.\hskip 1em plus 0.5em minus
  0.4em\relax IEEE, 2013, pp. 81--84.

\bibitem{koelstra2011deap}
S.~Koelstra, C.~Muhl, M.~Soleymani, J.-S. Lee, A.~Yazdani, T.~Ebrahimi, T.~Pun,
  A.~Nijholt, and I.~Patras, ``Deap: A database for emotion analysis; using
  physiological signals,'' \emph{IEEE transactions on affective computing},
  vol.~3, no.~1, pp. 18--31, 2011.

\bibitem{ni2021domain}
T.~Ni, Y.~Ni, J.~Xue, and S.~Wang, ``A domain adaptation sparse representation
  classifier for cross-domain electroencephalogram-based emotion
  classification,'' \emph{Frontiers in Psychology}, p. 3015, 2021.

\bibitem{tao2021multi}
J.~Tao and Y.~Dan, ``Multi-source co-adaptation for eeg-based emotion
  recognition by mining correlation information,'' \emph{Frontiers in
  Neuroscience}, p. 401, 2021.

\bibitem{pandey2021subject}
P.~Pandey and K.~Seeja, ``Subject independent emotion recognition system for
  people with facial deformity: an eeg based approach,'' \emph{Journal of
  Ambient Intelligence and Humanized Computing}, vol.~12, no.~2, pp.
  2311--2320, 2021.

\bibitem{kuang2021cross}
F.~Kuang, L.~Shu, H.~Hua, S.~Wu, L.~Zhang, X.~Xu, Y.~Liu, and M.~Jiang,
  ``Cross-subject and cross-device wearable eeg emotion recognition using
  frontal eeg under virtual reality scenes,'' in \emph{2021 IEEE International
  Conference on Bioinformatics and Biomedicine (BIBM)}.\hskip 1em plus 0.5em
  minus 0.4em\relax IEEE, 2021, pp. 3630--3637.

\bibitem{zheng2018emotionmeter}
W.-L. Zheng, W.~Liu, Y.~Lu, B.-L. Lu, and A.~Cichocki, ``Emotionmeter: A
  multimodal framework for recognizing human emotions,'' \emph{IEEE
  transactions on cybernetics}, vol.~49, no.~3, pp. 1110--1122, 2018.

\bibitem{liu2021comparing}
W.~Liu, J.-L. Qiu, W.-L. Zheng, and B.-L. Lu, ``Comparing recognition
  performance and robustness of multimodal deep learning models for multimodal
  emotion recognition,'' \emph{IEEE Transactions on Cognitive and Developmental
  Systems}, 2021.

\bibitem{li2019eeg}
Y.~Li, W.~Zheng, Z.~Cui, Y.~Zong, and S.~Ge, ``Eeg emotion recognition based on
  graph regularized sparse linear regression,'' \emph{Neural Processing
  Letters}, vol.~49, no.~2, pp. 555--571, 2019.

\bibitem{song2019mped}
T.~Song, W.~Zheng, C.~Lu, Y.~Zong, X.~Zhang, and Z.~Cui, ``Mped: A multi-modal
  physiological emotion database for discrete emotion recognition,'' \emph{IEEE
  Access}, vol.~7, pp. 12\,177--12\,191, 2019.

\bibitem{pedroni2019automagic}
A.~Pedroni, A.~Bahreini, and N.~Langer, ``Automagic: Standardized preprocessing
  of big eeg data,'' \emph{NeuroImage}, vol. 200, pp. 460--473, 2019.

\bibitem{croft2000eog}
R.~J. Croft and R.~J. Barry, ``Eog correction: which regression should we
  use?'' \emph{Psychophysiology}, vol.~37, no.~1, pp. 123--125, 2000.

\bibitem{parra2005recipes}
L.~C. Parra, C.~D. Spence, A.~D. Gerson, and P.~Sajda, ``Recipes for the linear
  analysis of eeg,'' \emph{Neuroimage}, vol.~28, no.~2, pp. 326--341, 2005.

\bibitem{pion2019iclabel}
L.~Pion-Tonachini, K.~Kreutz-Delgado, and S.~Makeig, ``Iclabel: An automated
  electroencephalographic independent component classifier, dataset, and
  website,'' \emph{NeuroImage}, vol. 198, pp. 181--197, 2019.

\bibitem{katz1988fractals}
M.~J. Katz, ``Fractals and the analysis of waveforms,'' \emph{Computers in
  biology and medicine}, vol.~18, no.~3, pp. 145--156, 1988.

\bibitem{jacob2019application}
J.~E. Jacob, G.~K. Nair, A.~Cherian, and T.~Iype, ``Application of fractal
  dimension for eeg based diagnosis of encephalopathy,'' \emph{Analog
  Integrated Circuits and Signal Processing}, vol. 100, no.~2, pp. 429--436,
  2019.

\bibitem{maaten2008visualizing}
L.~v.~d. Maaten and G.~Hinton, ``Visualizing data using t-sne,'' \emph{Journal
  of machine learning research}, vol.~9, no. Nov, pp. 2579--2605, 2008.

\bibitem{chen2021personal}
H.~Chen, S.~Sun, J.~Li, R.~Yu, N.~Li, X.~Li, and B.~Hu, ``Personal-zscore:
  Eliminating individual difference for eeg-based cross-subject emotion
  recognition,'' \emph{IEEE Transactions on Affective Computing}, 2021.

\bibitem{zhu2015cross}
J.-Y. Zhu, W.-L. Zheng, and B.-L. Lu, ``Cross-subject and cross-gender emotion
  classification from eeg,'' in \emph{World Congress on Medical Physics and
  Biomedical Engineering, June 7-12, 2015, Toronto, Canada}.\hskip 1em plus
  0.5em minus 0.4em\relax Springer, 2015, pp. 1188--1191.

\bibitem{waschke2021behavior}
L.~Waschke, N.~A. Kloosterman, J.~Obleser, and D.~D. Garrett, ``Behavior needs
  neural variability,'' \emph{Neuron}, vol. 109, no.~5, pp. 751--766, 2021.

\bibitem{li2018hierarchical}
J.~Li, Z.~Zhang, and H.~He, ``Hierarchical convolutional neural networks for
  eeg-based emotion recognition,'' \emph{Cognitive Computation}, vol.~10,
  no.~2, pp. 368--380, 2018.

\end{thebibliography}
	\clearpage
	
	\begin{appendices} 
	\section{comparison of differential entropy and Katz’s fractal dimension on different datasets}\label{appendixA}
	For different datasets(SEED, SEED-IV, SEED-V, RCLS, MPED), we extracted differential entropy(Diffen) and Katz’s fractal dimension(KatzFD) features, and Personal-Zscore method was applied on each session dataset to eliminate the impact of EEG individual differences. Then, two kinds of metric were used to evaluate the performance of two features on different datasets:          
	
	Amount of effective information, correlation feature analysis was used to judge the effectiveness of each dimension with condition of P$\le$0.05, and the metric was calculated as the ratio between the number of valid feature and total feature. 
	 
	Accuracy of emotion recognition, we employed 2-class emotion recognition task with Leave-One-Session-Out cross-validation on each dataset, and accuracies as the second metric.
	
	\begin{table*}[b]
		\centering
		\caption{THE BASIC RESULTS OF DIFFERENT DATASET}
		\label{tab8}
		\resizebox{\textwidth}{!}{%
			\begin{tabular}{lcccccccccccc}
				\hline
				\multirow{3}{*}{\textbf{DataSet}} &
				\multicolumn{6}{c}{\textbf{Feature correlation analysis}} &
				\multicolumn{6}{c}{\textbf{Emotion Recognition Accuracy}} \\ \cline{2-13} 
				&
				\multicolumn{3}{c}{\textbf{differential entropy}} &
				\multicolumn{3}{c|}{\textbf{Katz’s fractal dimension}} &
				\multicolumn{3}{c}{\textbf{differential entropy}} &
				\multicolumn{3}{c}{\textbf{Katz’s fractal dimension}} \\
				&
				\textbf{Neg-Neu} &
				\textbf{Pos-Neu} &
				\textbf{Pos-Neg} &
				\textbf{Neg-Neu} &
				\textbf{Pos-Neu} &
				\multicolumn{1}{c|}{\textbf{Pos-Neg}} &
				\textbf{Neg-Neu} &
				\textbf{Pos-Neu} &
				\textbf{Pos-Neg} &
				\textbf{Neg-Neu} &
				\textbf{Pos-Neu} &
				\textbf{Pos-Neg} \\ \hline
				\textbf{SEED} &
				74.57\% &
				81.31\% &
				83.21\% &
				75.48\% &
				91.51\% &
				\multicolumn{1}{c|}{90.42\%} &
				54.65\% &
				76.40\% &
				77.19\% &
				56.64\% &
				83.40\% &
				83.45\% \\
				\textbf{SEED-IV} &
				51.73\% &
				45.65\% &
				49.98\% &
				57.81\% &
				60.07\% &
				\multicolumn{1}{c|}{58.72\%} &
				57.78\% &
				53.76\% &
				61.63\% &
				66.41\% &
				56.67\% &
				63.43\% \\
				\textbf{SEED-V} &
				50.75\% &
				50.41\% &
				58.03\% &
				58.20\% &
				59.60\% &
				\multicolumn{1}{c|}{66.87\%} &
				57.81\% &
				63.30\% &
				65.10\% &
				54.25\% &
				65.99\% &
				70.16\% \\
				\textbf{MPED} &
				58.27\% &
				68.48\% &
				70.68\% &
				69.82\% &
				79.32\% &
				\multicolumn{1}{c|}{74.81\%} &
				60.21\% &
				63.51\% &
				63.49\% &
				62.10\% &
				67.63\% &
				64.80\% \\
				\textbf{RCLS} &
				51.52\% &
				55.04\% &
				45.55\% &
				70.02\% &
				77.52\% &
				\multicolumn{1}{c|}{74.12\%} &
				62.93\% &
				59.50\% &
				46.46\% &
				62.25\% &
				66.39\% &
				54.75\% \\
				\textbf{Average} &
				\textbf{57.37\%} &
				\textbf{60.18\%} &
				\textbf{61.49\%} &
				\textbf{66.27\%} &
				\textbf{73.60\%} &
				\multicolumn{1}{c|}{\textbf{72.99\%}} &
				\textbf{58.68\%} &
				\textbf{63.29\%} &
				\textbf{62.78\%} &
				\textbf{60.33\%} &
				\textbf{68.02\%} &
				\textbf{67.32\%} \\ \hline
			\end{tabular}%
		}
	\end{table*}
	
	After analysis of Table \ref{tab8}, we found the following points:
	
	(1) Comparison of different feature, the Katz’s fractal dimension was better than differential entropy from two metric on each dataset.
	
	(2) Comparison of different dataset, the effective information of SEED was better than the others, but the abnormal accuracy was occurred on the negative-neutral emotion recognition task.\cite{li2018hierarchical}
	
	(3) Comparison of different emotion recognition task, the performance on the negative-neutral emotion recognition task was bad than the others on the most of datasets, which indicated that there was less differences between negative state and neutral state.
	
	To analyze the reason why Katz’s fractal dimension better than differential entropy, we compared two features from the view of algorithm. $X$ and $x$ represent EEG signals and each singal point. 
	
	\subsubsection{differential entropy}
	
	\begin{equation}
	f_{X} (x)=\frac{1}{\sqrt{2\pi \sigma ^{2}} } e^{-\frac{(x-\mu ^{2} )}{2\sigma ^{2} } } 
	\end{equation}
	\begin{equation}
	Diffen(X)=\int f_{X}(x) dx =\frac{1}{2} log 2\pi e\sigma ^{2} 
	\end{equation}
	\begin{equation}
	variance(x) = log(\frac{\sum_{1}^{n} (x_{i} -\bar{x} )^{2} }{n})
	\end{equation}

	As shown in formula(5), some constants exist on the equation, which change the feature value, but they have not influence on the relationship between samples. Therefore, we eliminated the constants and found that differential entropy is equivalent to calculating the variance of the signal with the log operation.
	
	\subsubsection{Katz’s fractal dimension}
	
	\begin{equation}
	KatzFD=\frac{\log(L / a)}{\log(d / a)}=\frac{\log(n)}{\log(n)+\log(d / L)}
	\end{equation}
	
	\begin{equation}
	Inverse\_power(x)=\frac{1}{1+power(x)} 
	\end{equation}
	
	\begin{equation}
	power(x)=log(\sum_{1}^{n-1} (x_{i} -x_{i+1})^{2})
	\end{equation}
	
	Katz’s fractal dimension has a similar situation to differential entropy’s, and there also exists some constants on the formula(7), where $n$ is the singal length and $L$ is a value near $n$. we performed the same operation on differential entropy, and found that the core of Katz’s fractal dimension algorithm is calculating the power of signal.
	
	Comparing Katz’s fractal dimension and differential entropy from the view of energy, differential entropy is equivalent to calculating the sum of amplitude on the signal with mean reference, which considered EEG as pulse signal. However, Katz’s fractal dimension quantifies the amount of signal amplitude change, characterizing the energy changes induced by brain activity during the period of time. Considering the brain as an organic organization, we considered that Katz’s fractal dimension was more reasonable than differential entropy in terms of representing energy.
	
	\clearpage
	
	\section{The calculation of Sample Aggregation Coefficient}\label{appendixB}
	
	In our previous work, we combined t-SNE and k-means algorithm for the Calculation of Individual Aggregation Coefficient (CIAC) algorithm to quantify the individual aggregation phenomenon, which directly judged the two sub-clusters were close in individual cluster\cite{chen2021personal}. Therefore, the judgement of two sub-clusters as an improvement was applied in Calculation of Sample Aggregation Coefficient for the quantification of different sample aggregation phenomena, which described the sample aggregation degree for clusters(trial, emotion, session, subject). The details are as follows.
	 
	\begin{algorithm}[htb]
		\caption{Calculation of the Sample Aggregation Coefficient}  
		\label{algthm1}
		\begin{algorithmic}[1]
			\Require
			
			$D_{h}=\left \{d_{i}\right \}_{i=1}^{n}$ high dimensional original dataset, n classes. $d_{i}=\left \{v_{ij}\right \}_{j=1}^{m}$ original class sample dataset, m samples.
			
			\Ensure
			\State $D_{l}=$\Call{t-SNE}{$D_{h},n$}, reduce dimension.
			\State $Label=$\Call{k-means}{$D_{l}$} 
			\For{class index $i = 1 \to n$}
			\State $N=\left \{n_{i}\right \}_{i=1}^{m}$  Count the number of samples with same label.
			\State Sort N, and obtain the first and second largest numbers of samples $N_{1},N_{2}$.
			\State Calculate the $distance$ and $radius$ of two sample sub-clusters.
			\If {$distance \ge \Call{sum}{radius}$}
			\State $rate_{i}=\frac{N_{1}}{m}$
			\Else
			\State $rate_{i}=\frac{N_{1}+N_{2}}{m}$
			\EndIf
			\EndFor  
			\State$SAC=$\Call{mean}{$\left \{rate_{j}\right \}_{j=1}^{n}$}
			\State \Return $SAC$
		\end{algorithmic} 
	\end{algorithm}

	Taking the calculation of trial sample aggregation coefficients as an example, the dataset is reduced to 2 dimensions firstly. Then, the k-means is applied for sample clustering. In i-th trial cluster, the sample number of the largest and second largest sub-clusters are counted, the distance between the two sub-clusters and the radius of both sub-cluster are considered as indicators to judge the closeness of both clusters. If two sub-clusters are judge as close, the samples of sub-clusters are considered as aggregation, the ratio of aggregation sample number and total trial sample is the i-th trial’s sample cover rate. Finally, the trial sample aggregation coefficient of dataset is the mean of all trial’s sample cover rates.
	
	\clearpage
	
	\section{trial label estimation algorithm}\label{appendixC}
	Based on the characteristics of the EEG sample aggregation phenomena, in order to cope with the situation of unknown cluster labels, we combined two clustering algorithms, DBSCAN and k-means, to estimated the number and label of sample clusters. The specific algorithms are as follows.
	
		\begin{algorithm}[htb]
		\caption{The predicted of trial number and trial label}  
		\label{algthm5}
		\begin{algorithmic}[1]
			\Require
			$Data$ the unknown session dataset.
			\Ensure
			\State $dist=$\Call{distance}{$Data$}, calculation of reachable distance between each sample.
			\State $seq=$\Call{sort}{$dist$}, sort dist to ascending sequence.
			\State $seq=$\Call{divide}{$seq$,$percentage$}, get the first $p$\% distance sequence with $l$ length.
			\For{ index $i = 1 \to l$} 
			\State $radius=seq_{i}$ 
			\State $cluster\_number_{i}$=\Call{DBSCAN}{$Data,radius$},  perform DBSCAN with i-th reachable distance.
			\EndFor
			\State $trial\_number$, reject invalid cluster number and get the highest occurrences cluster number.
			\State $embed\_data$=\Call{t-SNE}{$Data$,$trial\_number$}
			\State $trial\_label$=\Call{k-means}{$embed\_data,trial_number$}
			\State \Return $trial\_number, trial\_label$
		\end{algorithmic} 
	\end{algorithm}

	Firstly, the number of sample clusters is estimated by DBSCAN. Because DBSCAN determines the reachability between samples by drawing circles based on the center of each sample, the number of clusters will be influenced by the radius. For an unknown dataset, we calculate the distance between each sample and get the reachable distance sequence in ascending order. When the radius is less than the minimum reachable distance, it cannot be clustered, and the number of clusters is 0. 
	
	\begin{figure}[h]
		\centering
		\includegraphics[width=\linewidth]{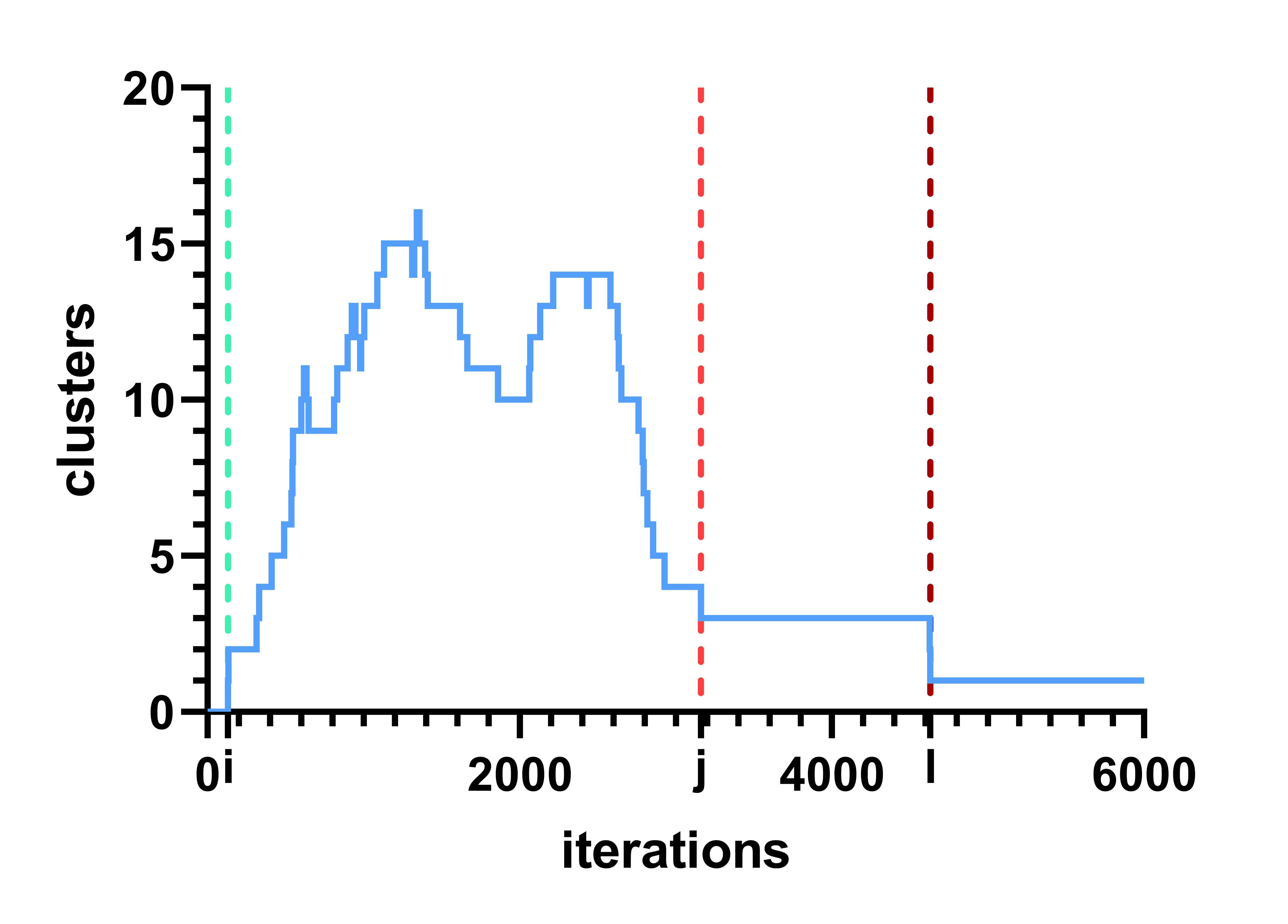}
		\caption{Changes in clustering number during the traversal of reachable distance sequence.}
		\label{fig12}
	\end{figure}
	
	Therefore, the reachable distance sequence is traversed, the tendency of clustering number with radius is shown in Fig. \ref{fig12}. As the radius increasing, the number of reachable samples also increasing, and the number of clusters will show a trend of growing and then decreasing until the radius is the $l$-th reachable distance, where all samples are reachable and the number of clusters is 1. Taking the sequence of clustering numbers from $i$ to $j$, $i$ is index that the clustering breakthrough at 0, $j$ is the index that clustering number down to a low and stable value, the occurrences for each clustering number are statistics, the clustering number with the highest occurrences as trial estimation. Specially, the stable results of 3 clusters between $j$ and $l$ indicate the emotional aggregation phenomenon. Finally, predicted trial labels are generated by k-means with trial estimation number. The algorithm can also be applied to session clustering estimation scenarios.

    \end{appendices}
\end{document}